\documentclass[12pt]{article}

\font\grande=cmr9.5 scaled \magstep4
\font\medio=cmr9.5 scaled \magstep2
\outer\def\beginsection#1\par{\medbreak\bigskip
      \message{#1}\leftline{\bf#1}\nobreak\medskip
\vskip-\parskip
      \noindent}
\usepackage{graphicx} 
\begin{document}
\bibliographystyle {unsrt}

\titlepage

\begin{flushright}
CERN-PH-TH/2012-133
\end{flushright}

\vspace{10mm}
\begin{center}
{\grande The symmetries of inflationary magnetogenesis}\\
\vspace{0.5cm}
{\grande and the plasma initial conditions}\\
\vspace{1.5cm}
 Massimo Giovannini
 \footnote{Electronic address: massimo.giovannini@cern.ch}\\
\vspace{1cm}
{{\sl Department of Physics, 
Theory Division, CERN, 1211 Geneva 23, Switzerland }}\\
\vspace{0.5cm}
{{\sl INFN, Section of Milan-Bicocca, 20126 Milan, Italy}}
\vspace*{0.5cm}
\end{center}

\vskip 0.5cm
\centerline{\medio  Abstract}
\vskip 0.2cm
As soon as the geometry expands quasi-exponentially the plasma sources are customarily tuned to zero as if the duration of the inflationary phase was immaterial for the gauge field fluctuations at large-scale. The serendipitous disappearance of the plasma (or even the partial neglect of its effects) depends on the symmetries of the system which are, in four space-time dimensions and in the simplest Abelian case, the invariance under Weyl rescaling and  the electromagnetic duality symmetry. The quantum, thermal and conducting initial conditions of inflationary magnetogenesis are classified and discussed with the aim of determining when plasma effects can be effectively disregarded. The speculative implications of a non-degenerate monopole plasma for the conservation of the large-scale electric flux are briefly examined.
\vskip 0.5cm

\noindent

\vspace{5mm}

\vfill
\newpage
\renewcommand{\theequation}{1.\arabic{equation}}
\setcounter{equation}{0}
\section{Introduction}
\label{sec1}
In four-dimensional curved space-times, besides the invariance under local gauge transformations, two symmetries are directly relevant to the gauge fields: the Weyl symmetry \cite{lich} and the electromagnetic duality symmetry \cite{duality1,duality2}. If the governing equations of a given quantum field are invariant under Weyl rescaling, the corresponding normal modes are not excited by the evolution of the geometry 
\cite{parker1,birrell,parker2} and particles are not produced. 
Duality rotates field strengths into their duals (i.e. tensors into pseudotensors) and it is therefore not an internal symmetry. 

The symmetries of the evolution equations of the gauge fields in curved space-times
are the main handle on the origin of the large-scale magnetism \cite{rev1,rev2,rev3}, a problem
dubbed magnetogenesis some time ago  \cite{mgenesis} but whose origin is much older as a number 
of relatively ancient but still very inspiring monographs demonstrates \cite{b1,b2,b3}. 
In the traditional framework magnetogenesis can take place either 
during or after a conventional phase of inflationary expansion. For this reason inflationary and post-inflationary magnetogenesis have been regarded as two separate and mutually exclusive possibilities.  Amplified quantum mechanical fluctuations are associated with inflationary magnetogenesis by implicitly disregarding the role of  electromagnetic sources. Conversely, post-inflationary magnetogenesis rests on the existence of appropriate currents operating inside the particle horizon (if present) at a given epoch in the life of the Universe. The purpose of this paper is to bridge the two aforementioned perspectives by scrutinizing the role of plasma sources in the situation where they are customarily neglected, i.e. during inflation. As it will be shown, the symmetries of inflationary magnetogenesis do not forbid the presence of electromagnetic sources unless they are fine-tuned to vanish initially for some theoretical prejudice. The present analysis completes and systematizes the results reported in  \cite{weyl} where a swift account of some of  the ideas contained in this paper has been illustrated. 

The duration of inflation is parametrized in terms of the total number of efolds\footnote{The number of efolds is the natural logarithm of the ratio between the scale factor at end (i.e. $a_{\mathrm{e}}$) and the beginning (i.e. $a_{\mathrm{b}}$) of the inflationary expansion. For the purposes of this introductory section we can conventionally agree that inflation starts as soon as an event horizon is formed.}, denoted by $N$. Since $N$ is not accessible to direct  
observations it is customary to introduce various (conceptually related) quantities such as $N_{\mathrm{min}}$ (i.e. the minimal number of efolds necessary to fix the problems of the standard big-bang cosmology) or $N_{\mathrm{max}}$ (i.e. the maximum value of inflationary efolds presently accessible to our cosmological observations). The value of $N_{\mathrm{min}}$ 
stems directly from the analysis of all the kinematical and dynamical problems of the standard hot big-bang scenario 
while the value of $N_{\mathrm{max}}$ can be determined by fitting the present size of the Hubble radius inside the event horizon of the quasi-de Sitter space-time.  The numerical value of $N_{\mathrm{max}}$ is not independent on the post-inflationary thermal history and may range from about $62$ (for the standard thermal history with sudden reheating) to 
larger figures if the reheating is delayed (see, for instance, \cite{m2,LL}). For the standard thermal history 
with sudden reheating $N_{\mathrm{min}} \simeq N_{\mathrm{max}}$. In the case of scalar and tensor modes of the geometry,
the conventional argument stipulates that when the total number of  efolds $N$ is much larger than $N_{\mathrm{max}}$ the effect of the initial conditions matters very little for any conclusion involving the amplification of quantum fluctuations. In the limit $N \gg N_{\mathrm{max}}$ the quantum fluctuations are automatically identified, in practice, with vacuum fluctuations. The problem of the initial conditions for the scalar and tensor fluctuations of the geometry has been discussed in the past; see, for instance, Refs. \cite{nonvac1,nonvac1a,nonvac1b} and \cite{nonvac2,nonvac3,nonvac3a,nonvac3b} for a still  incomplete list of articles.  

In contrast to what happens for the scalar and tensor modes of the geometry, it will be argued that the Weyl and duality 
symmetries determine if and when the initial conditions of the problem are to be treated classically or quantum mechanically.  Even conceding that the total duration of inflation is unknown, the inflationary phase is not eternal in the past either and potential sources cannot be neglected at the protoinflationary transition, i.e. 
when the background geometry starts accelerating. An exceedingly large number of efolds does not justify, per se,  the total or partial neglect of the electromagnetic sources during the inflationary evolution of the gauge fields. If the protoinflationary initial conditions are dominated by a relativistic plasma (containing either electric or magnetic sources) the whole system is invariant under Weyl rescaling. In the latter case the conductivity, may stay unsuppressed up to a critical efold (be it $N_{\mathrm{c}}$) which is not determined by the initial conditions but by the mass of the charge carriers. Charge carriers sufficiently light during inflation and with masses in the GeV range lead to $N_{\mathrm{c}} \sim {\mathcal O}(34)$ (see, for instance, section \ref{sec4}). In this example the sources can still be relevant $(N_{\mathrm{max}} - N_{\mathrm{c}})$ efolds before the end of inflation in contrast to the conventional set of assumptions. It is therefore not excluded that Weyl symmetry prevents the dissipation of protoinflationary Ohmic currents by so modifying the standard protocol for the assignment of the initial data in inflationary magnetogenesis.

In a realistic approach to the quantum and classical initial conditions of inflationary magnetogenesis it is safe 
 to enforce the global charge neutrality of the primeval plasma prior to the onset of inflation to avoid 
any phenomenological drawback \cite{LB}.  This 
requirement implies that the concentration of positively charged species is exactly balanced by the negatively charged ones. Denoting with $n_{+}$ and  $n_{-}$ the comoving concentrations of the species with positive and negative electric charges, we shall assume that $n_{+} = n_{-} = n_{0}$ where $n_{0}$ is the comoving value of the common charge concentration determining directly the Debye shielding length of the initial electric field. If $n_{0} \neq 0$ during the protoinflationary phase the evolution of the electric and of the magnetic fields is sharply different. The presence of an electric current breaks the duality symmetry which is instead unbroken in the absence of electromagnetic sources. The latter situation leads to the classical (or conducting) initial conditions.  Conversely, if the charge concentration vanishes, i.e. $n_{0} =0$, two complementary subclasses of initial conditions may arise depending on the duration of inflation. If the duration of inflation greatly exceeds $N_{\mathrm{max}}$ the initial conditions minimize, in practice, the Hamiltonian of the gauge field fluctuations. Finally if the total number of efolds is close to $N_{\mathrm{max}}$ (but always in the case $n_{0} =0$) the initial conditions for the evolution of the electric and magnetic fields are likely to be thermal.

The present paper is organized as follows. Section \ref{sec2} introduces the Weyl and duality symmetries in conformally 
flat background geometries.
In section \ref{sec3} the quantum mechanical treatment of the vacuum and of the thermal initial conditions 
is addressed when the total charge concentration vanishes (i.e. $n_{0} =0$). In section \ref{sec4} the plasma 
initial conditions, corresponding to the case $n_{0} \neq 0$, are examined. Section \ref{sec5} contains 
some considerations on a Lorentzian plasma of magnetic monopoles. 
The concluding remarks are collected in section \ref{sec6}.  To avoid lengthy digressions, 
some among the most relevant technical results have been collected in the appendix. 

\renewcommand{\theequation}{2.\arabic{equation}}
\setcounter{equation}{0}
\section{Weyl and duality symmetries}
\label{sec2}
The standard procedure for assigning the initial data to large-scale fluctuations 
stipulates, in a nutshell, that as soon as the background geometry inflates, all the sources present in the plasma can be serendipitously neglected with the sole exception of an (effective) cosmological term mainly produced by the potential of the inflaton. The previous recipe is justified when applied to the evolution of the geometry itself but it is arbitrary as far as the governing equations of the gauge fields are concerned. The point is that given a system of equations, there can be symmetries preventing the dissipation of the corresponding sources. 
To shed light on this aspect, the Weyl symmetry and the duality symmetry will now be swiftly scrutinized.

\subsection{Weyl symmetry}
Consider the following rescaling of the four-dimensional metric tensor\footnote{The signature of the metric is taken to be mostly 
minus. Four-dimensional indices are denoted by lowercase Greek letters while 
the spatial indices are denoted with lowercase Latin characters. Units $\hbar=c= \kappa_{\mathrm{B}} = 1$ will be assumed hereunder.}  
\begin{equation}
G_{\mu\nu} \to g_{\mu\nu} = q(x)\, G_{\mu\nu}, \qquad x= x^{\mu}=(\tau, x^{i}),
\label{WW1}
\end{equation}
where $x$ denotes the space-time coordinate. Under the transformation (\ref{WW1}) a generic Abelian vector field, the related field strength and its dual become:
\begin{equation}
 {\mathcal Y}^{\mu} \to Y^{\mu} = \frac{{\mathcal Y}^{\mu}}{q(x)},  \qquad 
{\mathcal Y}^{\mu\nu} \to Y^{\mu\nu} = \frac{{\mathcal Y}^{\mu\nu}}{q^2(x)},\qquad 
\tilde{{\mathcal Y}}^{\mu\nu} \to \tilde{Y}^{\mu\nu} = \frac{\tilde{{\mathcal Y}}^{\mu\nu}}{q^2(x)}, 
\label{WW2}
\end{equation}
where $Y_{\alpha\beta}$ and $\tilde{Y}^{\alpha\beta}$ are, respectively, the field strength and its four-dimensional dual defined with respect to the metric $g_{\mu\nu}$:
\begin{equation}
Y_{\alpha\beta} = \nabla_{[\alpha}Y_{\beta]} = \partial_{[\alpha} Y_{\beta]},\qquad 
\tilde{Y}^{\alpha\beta} = \frac{\epsilon^{\alpha\beta\rho\sigma}}{2 \sqrt{-g}} Y_{\rho\sigma};
\label{Q2} 
\end{equation}
$\nabla_{\alpha}$ is the covariant derivative with respect to the metric $g_{\mu\nu}$; ${\mathcal Y}_{\alpha\beta}$ and $\tilde{{\mathcal Y}}^{\alpha\beta}$ are the field strength and its dual but defined in terms of the metric $G_{\mu\nu}$.
Using Eqs. (\ref{WW1}) and (\ref{WW2})  it follows, in four space-time dimensions, that  
the vector potential transforms as:
\begin{equation}
 \sqrt{-G} {\mathcal Y}^{\mu} \to  \sqrt{-g} Y^{\mu} = q(x) \, \sqrt{- G} {\mathcal Y}^{\mu},
\label{WW4a}
\end{equation}
while the corresponding field strengths transform as:
\begin{equation}
\sqrt{-G} \,\, {\mathcal Y}^{\mu\nu} \to  \sqrt{-g} Y^{\mu\nu} = \sqrt{-G}\,\, {\mathcal Y}^{\mu\nu}, \qquad 
\sqrt{-G} \, \tilde{{\mathcal Y}}^{\mu\nu} \to  \sqrt{-g} \,\,\tilde{Y}^{\mu\nu} = \sqrt{-G} \,\, \tilde{{\mathcal Y}}^{\mu\nu}.
\label{WW4b}
\end{equation}
Equation (\ref{WW4b}) implies that  $\sqrt{-g} \, Y^{\mu\nu}$ and $\sqrt{-g}\, \tilde{Y}^{\mu\nu}$ are both invariant under Weyl rescaling. The generally covariant four-divergence of the field strength and  of its dual 
vanish in the vacuum (i.e. 
$\nabla_{\mu} \,Y^{\mu\nu} = \nabla_{\mu} \,\tilde{Y}^{\mu\nu}=0$) as a consequence of the Maxwell's equations that can also be 
expressed as $\partial_{\mu}( \sqrt{-g} \, Y^{\mu\nu})= \partial_{\mu}(\sqrt{-g} \, \tilde{Y}^{\mu\nu}) =0$, provided
 $\mathrm{det}(\,g_{\mu\nu}\,) \neq 0$. The previous chain of arguments proves the invariance of the original equations under Weyl rescaling thanks to Eq. (\ref{WW4b}). The generally covariant 
Lorentz gauge condition (i.e. $\nabla_{\mu} Y^{\mu} =0$) 
is not Weyl invariant as it follows from Eq. (\ref{WW4a}). Consequently, the Coulomb gauge turns out to be more suitable in curved space-times. This observation will be used in section \ref{sec3} and in appendix \ref{APPA}. 

Consider now, for sake of simplicity, the physical situation where $G_{\mu\nu} = \eta_{\mu\nu}$ 
($\eta_{\mu\nu}= \mathrm{diag}(1,\, -1,\, -1,\, -1)$ denotes the Minkowski metric). Such a situation is qualified as physical since, 
whenever $q(x)$ does not depend on the spatial coordinates but only upon the conformal time coordinate $\tau$, $g_{\mu\nu}$ describes, according to Eq. (\ref{WW1}), a Friedmann-Robertson-Walker geometry with flat spatial sections. 
The simultaneous presence of  gravitating electric and magnetic sources together with an effective 
cosmological term implies that the dynamics of $g_{\mu\nu}$ is dictated  
by the Einstein equations appropriately written in their contracted form:
\begin{equation}
R_{\mu \nu} = 8 \pi G \biggl[ T_{\mu\nu} - g_{\mu\nu} \,\frac{T}{2} \biggr],
\label{Q1a}
\end{equation}
where $R_{\mu\nu}$ is the Ricci tensor computed from the metric $g_{\mu\nu}$ while  $T_{\mu\nu}$ denotes the corresponding (total) energy-momentum 
tensor given as the sum of the energy momentum tensors of the sources (i.e. $t_{\mu\nu}$),  of the electromagnetic field (i.e. ${\mathcal T}_{\mu\nu}$) and of the (effective) cosmological 
constant $\Lambda$:
\begin{equation}
 T_{\mu\nu} = t_{\mu\nu} + {\mathcal T}_{\mu\nu} + \frac{\Lambda}{8\pi G} g_{\mu\nu}.
\label{Q1c}
\end{equation}
Since ${\mathcal T}_{\mu\nu}$ is traceless, only the two remaining terms of Eq. (\ref{Q1c}) will contribute 
to $T= T_{\mu}^{\mu}$ in Eq. (\ref{Q1a}). The Ricci tensor, appearing at the left hand side of Eq. (\ref{Q1a}) reads, in explicit terms:
\begin{equation}
R_{\mu\nu} = \frac{1}{2}\biggl\{-\biggl[  \partial_{\alpha}Q \partial^{\alpha} Q + \partial_{\alpha} \partial^{\alpha} Q\biggr]\eta_{\mu\nu} +  \partial_{\mu}\partial_{\nu} Q - 2 \partial_{\mu}Q \partial_{\nu} Q\biggr\},
\label{Q2a}
\end{equation}
where $Q = \ln{q}$. The scaling of Eq. (\ref{Q2a}) with $Q$ must be compared with the energy-momentum tensor appearing at the right hand side of Eq. (\ref{Q1a}) whose different contributions scale as follows:
\begin{equation}
T_{\mu\nu} =  e^{- \delta\, Q} \overline{t}_{\mu\nu} + 
e^{- Q} \,\overline{{\mathcal T}}_{\mu\nu} + \frac{\Lambda}{8\pi G}  e^{Q}\,\eta_{\mu\nu}.
\label{Q2aa}
\end{equation}
In Eq. (\ref{Q2aa})  $\overline{t}_{\mu\nu}$ and  $\overline{{\mathcal T}}_{\mu\nu}$ denote 
the rescaled energy-momentum tensors of the sources and of the electromagnetic field. Since under
Weyl rescaling both $\sqrt{- g} Y^{\mu\nu}$ and $\sqrt{-g} \tilde{Y}^{\mu\nu}$ are invariant, 
${\mathcal T}_{\mu\nu}$ must scale as $q^{-1} = e^{-Q}$. 
Relativistic (or non-relativistic) charge carriers lead to different values of $\delta$ in Eq. (\ref{Q2aa}); 
quite generically, however, $\delta > 0$. In summary Eqs.  (\ref{Q2a}) and (\ref{Q2aa}) show that Eq. (\ref{Q1a}) is not Weyl-invariant. Furthermore the scaling properties of the sources at the right hand side of Eq. (\ref{Q1a})  guarantee that as soon as $Q$ increases, the effective cosmological constant dominates, at least asymptotically.

The evolution equations of the gauge fields in the presence of electric and magnetic currents are, respectively
\begin{equation}
\nabla_{\mu} Y^{\mu\nu} = 4 \pi j^{\nu}, \qquad \nabla_{\mu} \tilde{Y}^{\mu\nu} = 4 \pi \ell^{\nu},
\label{Q1b}
\end{equation}
we do not consider here dyons\footnote{Dyons are idealized point particles carrying both magnetic and electric charges; in this paper we shall confine our attention to the case of electric currents since this is, after all, the only realistic case.
The magnetic currents are briefly discussed in section \ref{sec5}.} but  either a  
plasma of electric charges or a plasma of magnetic charges. If $\ell^{\nu} =0$ and  $j^{\nu}\neq 0$ the plasma is purely electric since $j^{\nu}$ is a current of electric charges. If $\ell^{\nu} \neq 0$ and $j^{\nu} =0$ the plasma is purely magnetic since $\ell^{\nu}$ is a current of magnetic monopoles.
As established from the general discussion of Eqs. (\ref{WW4a})--(\ref{WW4b}), Eqs.  (\ref{Q1b}) and (\ref{Q1c})  are Weyl-invariant provided $\sqrt{-g} \, j^{\nu}$ or $\sqrt{-g} \ell^{\nu}$  are separately Weyl-invariant. 
Since we do not have any evidence of magnetic monopoles the electric plasma must be regarded as physically more realistic from the viewpoint of protoinflationary initial conditions. In the case of a purely electric plasma characterized by an Ohmic current $j^{\nu}$ Eq. (\ref{Q1b}) 
can be written as:
\begin{equation} 
\partial_{\mu} \biggl[ \sqrt{ - g}\,  Y^{\mu\nu} \biggl] = 4 \pi \sqrt{- g} \, \sigma(x) \, Y^{\nu\alpha}\, u_{\alpha}, 
\label{Q3}
\end{equation}
together with the supplementary conditions $\partial_{\mu} ( \sqrt{ - g}\,  \tilde{Y}^{\mu\nu} )=0$  and $g^{\alpha\beta} \, u_{\alpha} \, u_{\beta} =1$. Equation (\ref{Q3}) is invariant under Weyl rescaling provided
\begin{equation}
\sigma(x) \to \overline{\sigma}(x) = \sqrt{q(x)} \sigma(x), \qquad u_{\alpha}(x) \to \overline{u}_{\alpha}(x) = \frac{u_{\alpha}(x)}{\sqrt{q(x)}}.
\label{Q4}
\end{equation}
The scaling law of the conductivity reported in Eq. (\ref{Q4}) holds, for instance,  in the case of a relativistic plasma in approximate thermal equilibrium when the temperature exceeds both the mass of the charge carriers and the chemical potentials. This conclusion is well known \cite{lich} and has been used in a number of interesting 
analyses \cite{w1,w2,w3,w4} but all related to the post-inflationary evolution of the electromagnetic fields. 
We notice here that the conclusions of Eqs. (\ref{Q3}) and (\ref{Q4}) are qualitatively different from the ones 
deduced from Eqs. (\ref{Q2a}) and (\ref{Q2aa}): while asymptotically for large $Q$ the electromagnetic sources
are exponentially suppressed in the Einstein equations, they are not suppressed in Eq. (\ref{Q3}). Even if 
the electromagnetic sources can be ignored in the asymptotic dynamics of the geometry 
their presence cannot be ignored when assigning the initial data for the gauge fields.

The symmetric argument can be repeated for a purely magnetic plasma 
(i.e. $\ell^{\nu}\neq 0$ and $j^{\nu} =0$) where Eq. (\ref{Q1b}) becomes:
\begin{equation}
\partial_{\mu} \biggl[ \sqrt{ - g}\,  \tilde{Z}^{\mu\nu} \biggl] = 4 \pi \sqrt{- g} \, \sigma_{\mathrm{m}} 
\tilde{Z}^{\nu\alpha}\, u_{\alpha}, 
\label{Q5}
\end{equation}
together with the conditions $\partial_{\mu} (\sqrt{ - g}\,  Z^{\mu\nu})=0$ and  $g^{\alpha\beta} \, u_{\alpha} \, u_{\beta} =1$; in Eq. (\ref{Q5}) $\sigma_{\mathrm{m}}$ denotes the magnetic conductivity.
As in the case of Eqs. (\ref{Q3}), $ \sqrt{ - g}\,  Z^{\mu\nu} $ and 
$\sqrt{ - g}\,  \tilde{Z}^{\mu\nu}$ are separately invariant under Weyl rescaling. The right hand side of Eq. (\ref{Q5}) 
is Weyl invariant provided $\sigma_{\mathrm{m}}$ and $u_{\alpha}$ scale, respectively, as $\sqrt{q(x)}$ and 
$1/\sqrt{q(x)}$. A magnetic plasma is physically very different from the case 
of an electric plasma. The passage from a purely electric to a purely magnetic plasma is an example of duality rotation 
which will be introduced  hereunder. The monopole plasma will be briefly discussed in section \ref{sec5}
since it is per se interesting but not strictly central to the main theme of this paper. 
 
The main lesson to be drawn is, in short, the following. The results of Eqs. (\ref{Q3}), (\ref{Q4}) and (\ref{Q5}) provide two examples where the Weyl symmetry prevents the damping of the electromagnetic sources.  The sources cannot be neglected in spite of the duration of inflation unless they were already absent at the beginning or unless 
they are fine-tuned to zero by fiat. Even more concretely, if the temperature of the charge carriers is approximately comparable with the radiation temperature (as it happens when the plasma parameter is very small, see section \ref{sec4}) the whole plasma is Weyl invariant in the relativistic regime when the masses of the charge carriers and the chemical potentials can be neglected.  

It is appropriate to mention, as a side remark, that some of the considerations of the present section can be formulated in terms of an antisymmetric six-tensor and a six-current as suggested originally by Dirac \cite{dirac}. The latter approach does not add any immediate advantage for the physical considerations of this analysis but has the virtue of enlightening the connections between Weyl spaces and the special conformal transformations \cite{barut,rorlich}. These 
interesting connections will be left aside; we shall instead proceed with our four-dimensional formalism without grouping the field strengths and their duals in a (constrained) six-dimensional antisymmetric tensor field.

\subsection{Duality symmetry}
Consider a rotation of the field strength and of its dual parametrized by an angle $\vartheta$:
\begin{equation}
Y^{\mu\nu}\to Z^{\mu\nu} = \cos{\vartheta}\, Y^{\mu\nu} + \sin{\vartheta} \, \tilde{Y}^{\mu\nu}, \qquad 
\tilde{Y}^{\mu\nu} \to \tilde{Z}^{\mu\nu} = - \sin{\vartheta} \, Y^{\mu\nu} + \cos{\vartheta} \, \tilde{Y}^{\mu\nu}.
\label{Q3a}
\end{equation}
Under Eq. (\ref{Q3a}) Eq. (\ref{Q1b}) transforms as:
\begin{equation}
\nabla_{\mu} \tilde{Z}^{\mu\nu} = 4 \pi L^{\nu}, \qquad \nabla_{\mu} Z^{\mu\nu} = 4 \pi J^{\nu},
\label{Q4a}
\end{equation}
provided 
\begin{equation}
j^{\nu} \to J^{\nu} = \cos{\vartheta}\, j^{\nu} + \sin{\vartheta} \,\ell^{\nu},\qquad \ell^{\nu} \to L^{\nu} = - \sin{\vartheta}\, j^{\nu} + \cos{\vartheta} \,\ell^{\nu}.
\label{Q5a}
\end{equation}
The duality transformation mapping an electric plasma into a magnetic plasma is realized for $\vartheta= \pi/2$ when  $Y^{\mu\nu} \to Z^{\mu\nu} = \tilde{Y}^{\mu\nu}$, $\tilde{Y}^{\mu\nu} \to \tilde{Z}^{\mu\nu}=  - Y^{\mu\nu}$, $j^{\nu} \to J^{\nu} =\ell^{\nu}$ and $\ell^{\nu} \to L^{\nu} = - j^{\nu}$. In the latter case Eqs. (\ref{Q4}) and (\ref{Q5}) can be explicitly related. 

The duality symmetry can be generalized to the case when the kinetic term of the gauge fields is coupled to a scalar degree of freedom $\lambda(x)$. Consider then the following action: 
\begin{equation}
S_{Y} = - \frac{1}{16\pi} \int d^{4} x \, \sqrt{- g}\, \lambda(x) \, Y_{\alpha\beta} \,Y^{\alpha\beta}.
\label{act1}
\end{equation}
Recalling that $Y_{\mu} = (Y_{0}, \, - Y_{i})$, in the Coulomb gauge (i.e. $Y_{0} =0$  and $\vec{\nabla}\cdot \vec{Y} =0$), we shall have 
\begin{equation}
Y_{0i} = a^2 \, e_{i} = - \partial_{\tau} Y_{i}, \qquad Y_{ij} = - a^2 \epsilon_{i\,j\,k} b^{k} = \partial_{i} Y_{j} - \partial_{j} Y_{i}. 
\label{eb}
\end{equation}
The electric and the magnetic fields $\vec{e}$ and $\vec{b}$ are related to the canonical fields  
$\vec{E}$ and $\vec{B}$ as:
\begin{equation}
\vec{E} = \sqrt{\frac{\lambda}{4 \pi}} a^2 \, \vec{e} = - \vec{y}^{\, \prime} + {\mathcal F}\,\vec{y},\qquad 
\vec{B} =   \sqrt{\frac{\lambda}{4 \pi}} a^2 \, \vec{b} = \vec{\nabla}\times \vec{y},
\label{Tmn3}
\end{equation}
where ${\mathcal F} = \sqrt{\lambda}^{\,\prime}/\sqrt{\lambda}$ is the rate of variation of $\sqrt{\lambda}$.
The consistent use of $\vec{B}$, $\vec{E}$ and $\vec{y}$ avoids potential problems in the identification 
of the electric and magnetic degrees of freedom. The canonical fields appearing in  Eq. (\ref{I7}) are exactly
the normal modes of Eq. (\ref{Tmn3}). In this source-free case the analog of Eq. (\ref{Q1b}) can be written in terms of the canonical electric and magnetic fields:
\begin{equation}
\vec{\nabla} \times( \sqrt{\lambda} \vec{B}) - \frac{\partial}{\partial \tau} ( \sqrt{\lambda} \vec{E}) =0, \qquad 
\frac{\partial}{\partial \tau}\biggl( \frac{\vec{B}}{\sqrt{\lambda}}\biggr) + \vec{\nabla} 
\times \biggl(\frac{\vec{E}}{\sqrt{\lambda}}\biggr) =0.
\label{I7}
\end{equation}
The system (\ref{I7}) is left invariant by the generalized duality transformation: 
\begin{equation}
\vec{E} \to - \vec{B},\qquad \vec{B} \to \vec{E},\qquad \sqrt{\lambda} \to \frac{1}{\sqrt{\lambda}}.
\label{I8} 
\end{equation} 
The symmetry of Eq. (\ref{I8}) holds in the absence of electromagnetic sources. The separate addition of either an electric current or of a monopole current breaks the duality symmetry. 

In general terms $\lambda= \lambda[\varphi(x),\, \psi(x),\,...]$ may be a functional of various scalar degrees of freedom present in the model such as the inflaton $\varphi$ \cite{DT1}, the dilaton \cite{DT2},
a dynamic gauge coupling \cite{DT3, DT5} (see also \cite{DT5a,DT5b} and \cite{mgenesis}). The field $\lambda$ can be a functional of a 
spectator field $\psi$, \cite{DT6,DT6a} (see also \cite{DT7a,DT7b}) evolving during the inflationary phase; in this case there is 
no connection between the evolution of $\lambda$ and the gauge coupling. 
The physical features of the various models are different: while some of these ideas are preferentially realized in the case 
of bouncing models \cite{mgenesis,DT2}, some other are compatible with the 
standard inflationary paradigm \cite{DT1,DT3,DT5,DT6}. 

In the present investigation we shall preferentially consider models compatible with the conventional inflationary scenario where $\lambda$  depends on a spectator field. It is also possible 
to think in terms of an effective gauge coupling but, in this case, it is necessary to assume that the gauge coupling is stronger at the beginning, then it decreases and so that $\sqrt{\lambda}$ is bound to increase. The latter requirement fits perfectly with 
the observation that in the conventional inflationary expansion the gravitational coupling is strong: the curvature scale at which inflation takes place is  ${\mathcal O}(10^{-5} \, M_{\mathrm{P}})$ and, as we go backward in the past, the curvature gets even stronger. In this situation the electric fields are always deamplified and potential backreaction effects are negligible as discussed within different perspectives in \cite{DT3,DT5,DT6}. If the 
gravitational coupling is weak at the beginning (i.e. the curvature is vanishing small and the initial conditions 
are given in flat-space time) it is natural to ask that the gauge coupling is also small at the beginning and this 
happens, for instance, in the case of bouncing models \cite{DT2,bounce1,bounce2}. 

When the initial conditions 
are dominated by a globally neutral plasma the situation is yet different insofar as the correct expansion parameter which guarantees the validity of the plasma approximation is the plasma parameter (i.e. the inverse of the number of particles present in the Debye sphere) and not simply the gauge coupling constant. The plasma parameter is bound 
to decrease during the inflationary phase since this evolution guarantees that  
the Debye shielding scale gets larger, the electric fields are screened and, in this way, they lead to irrelevant 
backreaction effects throughout the whole inflationary stage. 
\renewcommand{\theequation}{3.\arabic{equation}}
\setcounter{equation}{0}
\section{Thermal and quantum initial condition}
\label{sec3}
In this section it will be assumed that the concentration of the positively and negatively charged species is fine tuned to vanish even if, 
according to the considerations on the Weyl and duality symmetries reported in section \ref{sec2}, this is not the most generic situation. It is though interesting to see what happens when the conventional quasi-de Sitter phase is preceded by a decelerated stage of expansion
\begin{equation}
\lim_{H_{*}\, t \ll 1}\, \dot{a}(t) = \lim_{H_{*}\, t \gg 1}\, \dot{a}(t) \, \geq \,0,\qquad \lim_{H_{*}\, t \ll 1} \ddot{a}(t) \,< \,0, \qquad  \lim_{H_{*}\, t \gg 1}\, \ddot{a}(t) \, \geq \,0,
\label{I1}
\end{equation}
where $H_{*}$ defines the time-scale of the protoinflationary transition and $a(t)$ is the scale factor in cosmic time. According to Eq. (\ref{I1}), the acceleration changes its sign as soon as the inflationary event horizon is formed while the background expands both before and after the transition. 
The background geometry is conformally flat (i.e. $g_{\mu\nu} = a^2(\tau) \eta_{\mu\nu}$) and falls into the more general class of 
examples analyzed in section \ref{sec2}. The connection between the cosmic time coordinate $t$ and the conformal time coordinate $\tau$ is given, as usual, by $a(\tau) \, d\tau = d t$.  To avoid potential misunderstandings, 
it should be clear that the transition from a deceleration to acceleration has nothing to do with the so-called bouncing behaviour where the background passes from contraction to expansion or vice versa (see, e.g. \cite{bounce1,bounce2} and references therein). 
\subsection{Protoinflationary initial data}
In what follows, the protoinflationary sources are supposed to be in thermal equilibrium at a physical temperature $\overline{T}_{r}$ such that\footnote{Temperature-dependent phase transitions \cite{nonvac1,nonvac1a,nonvac1b,nonvac2,nonvac3,nonvac3a,nonvac3b} lead to an initial thermal state 
for the metric perturbations. The same kind of approach will be pursued in the present section. If the initial state is not thermal (but it is not the vacuum either) the present formalism applies with the difference that the spectral dependence of the mean number of photons will not necessarily have a Bose-Einstein form.}: 
\begin{equation}
H_{r}^2 M_{\mathrm{P}}^2 = \frac{8 \pi}{3} \overline{\rho}_{r}, \qquad \overline{\rho}_{r} = \frac{\pi^2}{30} g_{\mathrm{th}} 
\, \overline{T}_{r}^4,
\label{rate1}
\end{equation}
where $g_{\mathrm{th}}$ is the effective number of spin degrees of freedom. For $H_{*} \leq H_{r}$ the universe 
inflates driven, for instance, by a single inflaton field so that the inflationary curvature scale is determined as
\begin{equation}
\xi^2 = \biggl(\frac{H}{M_{\mathrm{P}}}\biggr)^2 = \frac{8\pi}{3} \biggl(\frac{V}{M_{\mathrm{P}}^4} \biggr),
\label{rate2}
\end{equation}
where $V$ denotes the inflaton potential and $\xi$ is fixed by the amplitude of the power spectrum of curvature perturbations ${\mathcal A}_{{\mathcal R}}$ at the pivot scale $k_{\mathrm{p}} =0.002\, \mathrm{Mpc}^{-1}$. In terms 
of the WMAP data alone \cite{wmap} and, in particular,  for the  7yr data release \cite{wmap7} analyzed in the light of the vanilla $\Lambda$CDM paradigm\footnote{We remind that in the acronym $\Lambda$CDM the $\Lambda$ stands for the 
late-time dark energy component and CDM stands for the late time cold dark matter component. The early completion 
of the $\Lambda$CDM paradigm contemplates a conventional inflationary phase with standard 
post-inflationary thermal hystory.} we have: 
\begin{equation}
\xi = \sqrt{\pi \, \epsilon\, {\mathcal A}_{{\mathcal R}}}, \qquad {\mathcal A}_{\mathcal R} =  (2.43\pm 0.11)\times 10^{-9},
\label{rate3} 
\end{equation}
where $\epsilon = - \dot{H}/H^2$ is the slow roll parameter.  The present value of $\overline{T}_{r}$ is 
\begin{equation}
\overline{T}_{r}(t_{0}) = Q_{i} \, \overline{T}_{\mathrm{max}} \, \biggl(\frac{H_{\mathrm{rh}}}{H}\biggr)^{\alpha - 1/2} \, e^{- N}\, 
\biggl(\frac{2 \Omega_{\mathrm{R}0}}{\pi \,\epsilon\, {\mathcal A}_{{\mathcal R}}} \biggr)^{1/4} \, \sqrt{\frac{H_{0}}{M_{\mathrm{P}}}},
\label{rate4}
\end{equation}
with 
\begin{equation}
Q_{i} = \frac{\overline{T}_{r}(t_{r})}{\overline{T}_{\mathrm{max}}}, \qquad \overline{T}_{\mathrm{max}} = \biggl( 
\frac{45}{4\pi^3 g_{\mathrm{th}}}\biggr)^{1/4} \sqrt{H \, M_{\mathrm{P}}}.
\label{rate5}
\end{equation}
Note that $Q_{i} \leq 1$ as long as $\overline{\rho}_{r}(t_{r}) \leq 3 H^2\, M_{\mathrm{P}}^2/(8 \pi)$.
The appropriate initial state is therefore a mixture characterized by a 
density matrix
\begin{equation}
\hat{\rho} = \sum_{\{n\}} P_{\{n \}} |\{n \}\rangle \langle \{n \}|,
\qquad P_{\{n\}} = \prod_{\vec{k}} \frac{\overline{N}_{k}^{n_{k}}}{( 1 + \overline{N}_{k})^{n_{k} + 1 }},
\label{matrix1}
\end{equation}
where $\overline{N}_{k}$ is the average occupation number of each Fourier mode and, following 
the standard notation, $ |\{n \}\rangle = |n_{\vec{k}_{1}} \rangle \otimes  |n_{\vec{k}_{2}} \rangle \otimes  |n_{\vec{k}_{3}} \rangle...$ where the ellipses stand for all the occupied modes of the field. 

\subsection{Two-point functions}
Recalling  the results of appendix \ref{APPA}, and, in particular, Eq. (\ref{dual8}),  
the canonical fields appearing in the duality-invariant Hamiltonian 
\begin{equation}
H_{Y}(\tau) = \frac{1}{2} \int d^3 x \biggl[ \vec{\pi}^{2} + 2 \,{\mathcal F} \, \vec{\pi} \cdot \vec{y} + 
\partial_{i} \vec{y} \cdot \partial^{i} \vec{y}\biggr],
\label{dualham}
\end{equation}
can be promoted to quantum operators (i.e. $y_{i} \to \hat{y}_{i}$ and $\pi_{i} \to \hat{\pi}_{i}$) 
obeying (equal time) commutation relations:
\begin{equation}
[\hat{y}_{i}(\vec{x}_{1},\tau),\hat{\pi}_{j}(\vec{x}_{2},\tau)] = i \Delta_{ij}(\vec{x}_{1} - \vec{x}_{2}),\qquad 
\Delta_{ij}(\vec{x}_{1} - \vec{x}_{2}) = \int \frac{d^{3}k}{(2\pi)^3} e^{i \vec{k} \cdot (\vec{x}_{1} - \vec{x}_2)} P_{ij}(k), 
\label{dual17}
\end{equation}
where $P_{ij}(\hat{k}) = (\delta_{ij} - k_{i} k_{j}/k^2)$. The function $\Delta_{ij}(\vec{x}_{1} - \vec{x}_{2})$ is the transverse generalization of the Dirac delta function  and such an extension is 
mandatory since $ \vec{\nabla} \cdot \vec{E} =0$ (because of the Gauss constraint) and $\vec{\nabla}\cdot \vec{y}=0$ (because of the gauge condition). The field operators will then become (see also Eqs. (\ref{dual9})--(\ref{dual10}) of appendix \ref{APPA})
\begin{eqnarray}
\hat{y}_{i}(\vec{x},\tau) = \int\frac{d^{3} k}{(2\pi)^{3/2}} \sum_{\alpha} e^{(\alpha)}_{i}(\hat{k}) \, 
\biggl[ f_{k}(\tau) \, \hat{a}_{\vec{k},\,\alpha} e^{- i \vec{k} \cdot\vec{x}} +  f_{k}^{*}(\tau) \, \hat{a}^{\dagger}_{\vec{k},\,\alpha} e^{ i \vec{k} \cdot\vec{x}}\biggr],
\label{dual18}\\
\hat{\pi}_{i}(\vec{x},\tau) = \int\frac{d^{3} k}{(2\pi)^{3/2}} \sum_{\alpha} e^{(\alpha)}_{i}(\hat{k}) \, 
\biggl[ g_{k}(\tau) \, \hat{a}_{\vec{k},\,\alpha} e^{- i \vec{k} \cdot\vec{x}} +  g_{k}^{*}(\tau) \, \hat{a}^{\dagger}_{\vec{k},\,\alpha} e^{ i \vec{k} \cdot\vec{x}}\biggr],
\label{dual19}
\end{eqnarray}
where, $e^{(\alpha)}_{i}(\hat{k})$ (with $\alpha =1, \,2$) are two mutually orthogonal unit vectors which are also orthogonal to 
$\hat{k}$;  the creation and annihilation operators appearing in Eqs. (\ref{dual18}) 
and (\ref{dual19}) obey $[\hat{a}_{\vec{k},\alpha}, \hat{a}^{\dagger}_{\vec{p}\,\beta} ] = \delta_{\alpha\beta} \delta^{(3)}(\vec{k} + \vec{p})$. Fom Eq. (\ref{dual13}) the mode functions  $f_{k}$ and $g_{k}$ obey:
\begin{equation}
f_{k}' = g_{k} + {\mathcal F} f_{k},\qquad 
g_{k}' = - k^2 f_{k} - {\mathcal F} g_{k},
\label{DD21}
\end{equation}
and satisfy the Wronskian normalization condition 
$f_{k}(\tau)g_{k}^{*}(\tau) - f_{k}^{*}(\tau) g_{k}(\tau) =i$ which follows from enforcing the canonical commutators 
between the field operators. The magnetic power spectra and the electric power spectra can be computed from 
the density matrix of Eq. (\ref{matrix1}) by explicitly evaluating the following pair of correlators of
field operators:
\begin{eqnarray}
{\mathcal G}_{ij}^{\mathrm{(B)}} (r,\tau) &=& \langle \hat{B}_{i}(\vec{x},\tau) \, \hat{B}_{j}( \vec{x} + \vec{r},\tau) \rangle = \mathrm{Tr} \bigl[ \hat{\rho}\,  \hat{B}_{i}(\vec{x},\tau) \, \hat{B}_{j}( \vec{x} + \vec{r},\tau)\bigr],
\label{matrix2}\\
{\mathcal G}_{ij}^{\mathrm{(E)}} (r,\tau) &=& \langle \hat{E}_{i}(\vec{x},\tau) \, \hat{E}_{j}( \vec{x} + \vec{r},\tau) \rangle = \mathrm{Tr} \bigl[ \hat{\rho}\,  \hat{E}_{i}(\vec{x},\tau) \, \hat{E}_{j}( \vec{x} + \vec{r},\tau)\bigr].
\label{matrix3}
\end{eqnarray}
Since most of the forthcoming formulas are separately valid for magnetic and electric correlators, we shall 
use the notation ${\mathcal G}_{ij}^{(X)}$ where it is understood that $X= \mathrm{B},\, \mathrm{E}$. 
The isotropic and time-dependent correlators  
${\mathcal G}_{ij}^{X}(r,\tau)$ are transverse and can therefore be represented by means of the 
standard fluid parametrization (see, e.g. \cite{yaglom,landau2,compress}): 
\begin{equation}
{\mathcal G}_{ij}^{(X)}(r,\tau) = {\mathcal G}^{(X)}_{\mathrm{T}}(r,\tau) \delta_{ij} + \frac{r_{i} \, r_{j}}{r^2} \,\biggl[ {\mathcal G}^{(X)}_{\mathrm{L}}(r,\tau) - {\mathcal G}^{(X)}_{\mathrm{T}}(r,\tau)\biggr],\qquad \frac{\partial {\mathcal G}^{(X)}_{\mathrm{L}}}{\partial r} + \frac{2}{r} \bigl({\mathcal G}^{(X)}_{\mathrm{L}} - 
{\mathcal G}^{(X)}_{\mathrm{T}}\bigr) =0,
\label{matrix4}
\end{equation}
where ${\mathcal G}^{(X)}_{\mathrm{T}}(r,\tau)$ and ${\mathcal G}^{(X)}_{\mathrm{L}}(r,\tau)$
denote, respectively, the transverse and the longitudinal parts of the either magnetic or electric correlators; 
since the electric and magnetic correlators are both divergenceless they must also be subjected, to the differential 
relation appearing in Eq. (\ref{matrix4}) (see also, in a related hydrodynamical context, Ref. \cite{landau2,compress}). Using
Eq. (\ref{matrix3}) the explicit expressions of the transverse and longitudinal components of the electric 
and magnetic correlators become:
\begin{equation}
{\mathcal G}^{(X)}_{\mathrm{L}}(r,\tau) = \int \, d \ln{k} \, P_{X}(k,\tau) \, q_{\mathrm{L}}(k r),
\qquad   {\mathcal G}^{(X)}_{\mathrm{T}}(r,\tau) = \int \, d \ln{k} \, P_{X}(k,\tau) \, q_{\mathrm{T}}(k r),
\label{matrix7}
\end{equation}
where $P_{X}(k,\tau)$ denotes either the magnetic or the electric power 
spectrum:
\begin{equation}
P_{\mathrm{B}}(k,\tau) = \frac{k^5}{2\pi^2} |f_{k}(\tau)|^2 \, ( 2 \overline{N}_{k} + 1),
\qquad 
P_{\mathrm{E}}(k,\tau) = \frac{k^3}{2\pi^2} |g_{k}(\tau)|^2\, ( 2 \overline{N}_{k} +1).
\label{matrix9}
\end{equation}
The functions $q_{\mathrm{T}}(k r)$ and $q_{\mathrm{L}}(k r)$ of Eq. (\ref{matrix7}) are given in terms of spherical Bessel 
functions \cite{abr,tric}: 
\begin{equation}
q_{\mathrm{T}}(k r) = j_{0}(k r) - \frac{j_{1}(k r)}{k r}, \qquad q_{\mathrm{L}}(k r) = j_{0}(k r) - \frac{j_{1}(k r)}{k r} + j_{2}(k r).
\label{matrix10}
\end{equation}
With the previous notations for the power spectra, ${\mathcal G}_{ij}^{(X)}(r,\tau)$ can be represented in Fourier space as
\begin{equation}
{\mathcal G}_{ij}^{(X)}(r,\tau) = \frac{1}{4\pi} \int \frac{d^{3} k}{ k^3} \, P_{X}(k,\tau) \, P_{ij}(\hat{k}) e^{- i \vec{k}\cdot \vec{r}}.
\label{matrix11}
\end{equation}
where $P_{ij}(\hat{k})$ is, again, the standard transverse projector.
\subsection{Power spectra from thermal initial conditions}
As discussed in Appendix \ref{APPB}, Eq. (\ref{DD21}) can be solved under different approximations. If 
$\sqrt{\lambda}$ evolves monotonically the magnetic and the electric power 
spectra of Eq. (\ref{matrix9}) can be parametrized as:
\begin{equation}
P_{X}(k,\tau, \tau_{\sigma}) = P_{X}(x,\,y,\,z) = \frac{k^4}{8 \pi \alpha^2} \, e^{- 2 z}\, {\mathcal M}_{X}(x, y) ( 2 \overline{N}_{k} + 1),
\label{SPP1}
\end{equation}
where $x(k, \tau_{\sigma})$, $y(k,\tau,\tau_{\sigma})$ and $z(\tau,\tau_{\sigma})$ are dimensionless variables:
\begin{equation}
 x(k,\tau_{\sigma}) = k \,\tau_{\sigma},\qquad y(k, \tau,\tau_{\sigma}) = \int_{\tau_{\sigma}}^{\tau} k\, \alpha(k,\tau') \, d\tau', \qquad 
z(\tau,\tau_{\sigma}) = 2 \pi \int_{\tau_{\sigma}}^{\tau}\sigma(\tau') \, d\tau'.
\label{app10}
\end{equation}
The functions ${\mathcal M}_{X}(x, y)$ with $X= \mathrm{B}$ and $X= \mathrm{E}$ are computed from the results 
of appendix \ref{APPB} and they are:
\begin{eqnarray}
 {\mathcal M}_{\mathrm{B}}(x, y) &=& \biggl\{ x |H^{(1)}_{\nu}(x)|^2 [ \alpha c(y) 
+ \Sigma s(y)]^2 + x  |H^{(1)}_{\nu-1}(x)|^2 s^2(y)
\nonumber\\
&-& x \, s(y) [ \alpha c(y) + \Sigma s(y)] \biggl[ H^{(1)}_{\nu}(x) H^{(2)}_{\nu-1}(x) + H^{(2)}_{\nu}(x)H^{(1)}_{\nu-1}(x)\biggr]\biggr\},
\label{DEF1}\\
{\mathcal M}_{\mathrm{E}}(x, y) &=& \biggl\{ x |H^{(1)}_{\nu-1}(x)|^2 [ \alpha c(y) 
- \Sigma s(y)]^2 + x |H^{(1)}_{\nu}(x)|^2  s^2(y)
\nonumber\\
&-& x \, s(y) [ \alpha s(y) - \Sigma c(y)] \biggl[ H^{(1)}_{\nu}(x) H^{(2)}_{\nu-1}(x) + H^{(2)}_{\nu}(x)H^{(1)}_{\nu-1}(x)\biggr]\biggr\},
\label{DEF2}
\end{eqnarray}
where $c(y) = \cosh{y}$, $s(y) = \sinh{y}$ and $\Sigma= 2\pi\sigma/k$.
The time $\tau_{\sigma}$ marks the beginning of the conducting phase at the end of inflation. In the sudden reheating approximation $\tau_{\sigma}$ 
coincides with the end of the inflationary phase. To evaluate explicitly the power spectra we must specify $\overline{N}_{k}$. 
The obvious choice is\footnote{The density 
matrix of Eq. (\ref{matrix1}) describes a mixed state obeying the Bose-Einstein statistics. A 
 Bose-Einstein multiplicity distribution is not sufficient to infer local thermal equilibrium. In various quantum optical situations (see, e.g. \cite{loudon2} pp. 159 and also \cite{mandel}) chaotic light (i.e. white light obeying as a Bose-Einstein distribution for each mode of the radiation field) 
is generated by a source in which atoms are kept at an excitation level higher than that in thermal equilibrium. In this sense the present calculation holds 
also in more general cases when $\overline{N}_{k}$ does not have a Bose-Einstein form.}
\begin{equation}
\overline{N}_{k} = \frac{1}{e^{k/k_{\mathrm{T}}} -1}, \qquad k_{\mathrm{T}} = T_{r} = a \, \overline{T}_{r},
\label{NN}
\end{equation}
where $T_{r}$ the comoving temperature of the initial thermal background. The power spectra (\ref{SPP1})  contain both the contributions from the vacuum fluctuations and from the thermal fluctuations. When $\overline{N}_{k} \ll1 $ (i.e. $k \gg k_{\mathrm{T}}$) the quantum fluctuations dominate whereas when $\overline{N}_{k} \gg 1$ (i.e. 
$k\ll k_{\mathrm{T}}$) the thermal fluctuations are the leading source of inhomogeneity in the initial conditions. In the limits $ x \ll 1$ and $\Sigma = 2 \pi \sigma/k \gg 1$,  
Eq. (\ref{SPP1}) (for $X = \mathrm{B}$ and $X= \mathrm{E}$) and Eqs. (\ref{DEF1})--(\ref{DEF2}) lead to the wanted power spectra for the magnetic and electric fields:
\begin{eqnarray}
&& P_{\mathrm{B}}(k,\tau) \simeq \frac{k^4}{8\pi} \, x |H^{(1)}_{\nu}(x)|^2\, e^{ - 2{\mathcal J}(k, \sigma)} \coth{\biggl[\frac{k}{2 k_{\mathrm{T}}}\biggr]}\biggl[ 1 + {\mathcal O}\biggl(\frac{1}{\Sigma}\biggr) \biggr],
\label{SPP3}\\
&& P_{\mathrm{E}}(k,\tau) \simeq \frac{k^4}{8\pi \Sigma^2} \, x |H^{(1)}_{\nu-1}(x)|^2\, e^{ -2 {\mathcal J}(k, \sigma)} \coth{\biggl[\frac{k}{2 k_{\mathrm{T}}}\biggr]}\biggl[ 1 + {\mathcal O}\biggl(\frac{1}{\Sigma}\biggr) \biggr],
\label{SPP4}
\end{eqnarray}
where ${\mathcal J}(k,\sigma)= [z(\tau,\,\tau_{\sigma})-y(k,\tau,\tau_{\sigma})]$ which becomes, using Eq. (\ref{app10}) and the results of appendix \ref{APPB}:
\begin{equation}
{\mathcal J}(k,\sigma) =  k \int_{\tau_{\sigma}}^{\tau} \biggl[ \Sigma(k,\sigma) - \sqrt{\Sigma^2(k,\sigma) -1} \biggr] d\tau'  \simeq  \int_{\tau_{\sigma}}^{\tau} \frac{k^2}{4\pi \sigma(\tau')} d\tau'.
\label{SPP5}
\end{equation}
It is useful to introduce the magnetic diffusivity scale $k_{\sigma}$ such that 
${\mathcal J}(k,\sigma) = k^2/k_{\sigma}^2$ where $k_{\sigma}^{-2}  = \int_{\tau_{\sigma}}^{\tau} \, d\tau' /[4\pi \sigma(\tau')]$. The evaluation 
of $k_{\sigma}$ is complicated by the fact that the integral of Eq. (\ref{SPP5}) extends well after $\tau_{\sigma}$. This estimate
can be made rather accurate by computing the transport coefficients of the plasma in different regimes \cite{CC1,CC2} (see also \cite{CC3}). 
For the present purposes, however, such an effort would be forlorn given the minuteness of the ratio $(k/k_{\sigma})^2$ for the 
phenomenologically interesting scales. By taking $\tau = \tau_{\mathrm{eq}}$ in Eq. (\ref{SPP5}) the result is 
\begin{equation}
\biggl(\frac{k}{k_{\sigma}}\biggr)^2 = \frac{4.75 \times 10^{-26}}{ \sqrt{2 \, h_{0}^2 \Omega_{\mathrm{M}0} (z_{\mathrm{eq}}+1)}} \, \biggl(\frac{k}{\mathrm{Mpc}^{-1}} \biggr)^2,
\label{CC4}
\end{equation}
where $\Omega_{\mathrm{M}0}$ is the present critical fraction in matter, $h_{0}$ is the Hubble rate in units of $100\, \mathrm{km}/(\mathrm{sec}\,\mathrm{Mpc})$ 
and $z_{\mathrm{eq}} + 1 = a_{0}/a_{\mathrm{eq}}\simeq 3200$ is the redshift of matter-radiation equality.  Eq. (\ref{CC4}) can be 
also used, with some caveats, to estimate $\Sigma$ or $k/\Sigma\simeq k^2/\sigma$. 
Equations (\ref{SPP3}) and (\ref{SPP4}) show that the electric power spectrum is suppressed not only inside the Hubble radius but also outside by the factor $\Sigma^{-2} \sim k^2/\sigma^2$.  The present value of the power spectrum can be estimated as
\begin{equation}
\frac{\sqrt{P_{{\mathcal B}}(k,\tau_{0})}}{\mathrm{Gauss}} = 10^{-10.84}\, 
\biggl(\frac{{\mathcal A}_{{\mathcal R}}}{2.43 \times 10^{-9}}\biggr)^{1/2}\,
\biggl(\frac{\Omega_{\mathrm{R}0}}{4.15 \times 10^{-5}}\biggr)^{1/2}\,
 \sqrt{f(\nu,k, \tau_{\sigma}, k_{\mathrm{T}}) },
\label{sol13}
\end{equation}
where $P_{\mathrm{B}}(k,\tau) = a^4(\tau) P_{{\mathcal B}}(k,\tau)$ and\footnote{The slope of the magnetic power spectra 
can be defined, from Eq. (\ref{sol14}), as $k^{n_{\mathrm{B}}-1}$ where $n_{\mathrm{B}} = (6 - 2 \nu)$. Note also that, in Eq. (\ref{sol14}), ${\mathcal K}(5/2) =1$. }
\begin{equation}
f(\nu,k, \tau_{\sigma}, k_{\mathrm{T}}) = {\mathcal K}(\nu)\, |k\tau_{\sigma}|^{5 - 2 \nu}\, \, \coth{\biggl[\frac{k}{2 k_{\mathrm{T}}}\biggr]}, 
\qquad  {\mathcal K}(\nu) = \frac{2 ^{2 \nu -1}}{9 \, \pi} \, \Gamma^2(\nu).
\label{sol14}
\end{equation}
The relevant scales of magnetogenesis (say between few tenths of Mpc and $10$ Mpc) have corresponding wavenumbers which are much larger than the putative 
effective temperature of the thermal background prior to the onset of inflation. 
The presence of the protoinflationary thermal background implies that 
\begin{equation}
\frac{k}{k_{\mathrm{T}}} = \frac{52.54}{Q_{i}}\, e^{- ( N_{\mathrm{max}} - N)}\, \biggl(\frac{k}{\mathrm{Mpc}^{-1}} \biggr)\,
\biggl(\frac{g_{\mathrm{th}}}{106.75}\biggr)^{1/4} \,\biggl(\frac{{\mathcal A}_{{\mathcal R}}}{2.43\times 10^{-9}}\biggr)^{1/4}
\biggl(\frac{\epsilon}{0.01}\biggr)^{1/4},
\label{KT1}
\end{equation}
where  $N_{\mathrm{max}}$ denotes the maximal number of efolds presently accessible by large-scale observations, i.e.
\begin{equation}
N_{\mathrm{max}} = 62.2 + \frac{1}{2} \ln{\biggl(\frac{\xi}{10^{-5}}\biggr)} - \ln{\biggl(\frac{h_{0}}{0.7}\biggr)}
+ \frac{1}{4} \ln{\biggl(\frac{h_{0}^2 \, \Omega_{\mathrm{R}0}}{4.15\times 10^{-5}}\biggr)}.
\label{KT2}
\end{equation}
For the numerical estimates it is useful to bear in mind that $\tau_{\sigma} = {\mathcal H}_{\sigma}^{-1} = 1/[a_{\sigma} H_{\sigma}]$; 
thus $k\tau_{\sigma}$ appearing in Eq. (\ref{sol14}) is expressible as:
\begin{equation}
\frac{k}{a_{\sigma} H_{\sigma}} = 3.22\times 10^{-25} \, \biggl(\frac{k}{\mathrm{Mpc}^{-1}}\biggr) \, \biggl(\frac{\epsilon}{0.01}\biggr)^{-1/4} \, 
\biggl(\frac{{\mathcal A}_{\mathcal R}}{2.43\times 10^{-9}} \biggr)^{-1/4}.
\label{ex}
\end{equation}
As already anticipated in section \ref{sec1} the value of $N_{\mathrm{max}}$ is close, in the sudden reheating approximation, to the minimal number of efolds $N_{\mathrm{min}}$ needed to solve the kinematic 
problems of the standard cosmological model (i.e. $N_{\mathrm{min}} \simeq N_{\mathrm{max}} \simeq N_{1}$). Recalling the fiducial set of 
cosmological parameters determined on the basis of the $\Lambda$CDM paradigm and in the light of the WMAP7 data \cite{wmap,wmap7}, Eq. (\ref{KT2}) gives $N_{1} = 63.6 + 0.25 \ln{\epsilon}$. By definition $N_{\mathrm{max}}$ is derived by requiring that the present 
size of the Hubble radius is all contained in the event horizon at the onset 
of the inflationary phase.
The figures given in Eq. (\ref{KT2}) are for guidance olnly:  $N_{\mathrm{max}}$ can be much larger if the 
reheating is delayed. If right after a conventional inflationary phase  the Universe expands at a rate 
which is slower than radiation $N_{\mathrm{max}}$ increases. In particular, if, after inflation, the 
energy density of the plasma is dominated by a stiff source with sound speed 
coinciding with the speed of light we get to the estimate $N_{\mathrm{max}} = 78.3 + (1/3) \ln{\epsilon}$
where it has been assumed that the stiff phase starts after inflation and stops right before big-bang nucleosynthesis
(see, e.g. \cite{lid} and references therein). 

In Eq. (\ref{KT1}) the value of  $Q_{i}$ can be, at most, ${\mathcal O}(1)$. This means that $k \gg  k_{\mathrm{T}}$ in the range of scales $0.1\, \mathrm{Mpc}^{-1} <\, k \, < \,10 \, \mathrm{Mpc}^{-1}$ and that the thermal correction is also, at most, ${\mathcal O}(1)$, since, in this range, $\coth{[k/(2 k_{\mathrm{T}})]} \simeq {\mathcal O}(1)$. The wavelength associated with the protoinflationary thermal background is then always larger than the typical length-scale 
related to the gravitational collapse of the protogalaxy, i.e. ${\mathcal O}(\mathrm{Mpc})$; such a scale is inside the horizon prior to matter-radiation 
equality. Thermal initial conditions are only relevant for $k \ll 0.1\, \mathrm{Mpc}^{-1}$. From Eq. (\ref{sol13}),  for $\nu =5/2$ and $N = N_{\mathrm{max}}$, $\sqrt{P_{\mathcal B}(k,\tau_{0})} \simeq 10^{-11} \, \mathrm{Gauss}$ in the range $k > 0.1\,\mathrm{Mpc}$ while 
$\sqrt{P_{\mathcal B}(k,\tau_{0})} \simeq 10^{-15} \, \mathrm{Gauss}$ for $k \sim 10^{-5} \mathrm{Mpc}^{-1}$ roughly corresponding 
to the present value of the Hubble radius. It is finally appropriate to note that for $\nu=1/2$  no amplification takes place (since $\lambda$ is constant) 
and the power spectrum will be ${\mathcal O}(10^{-61}) \, (k/\mathrm{Mpc}^{-1})^2\, \mathrm{Gauss}$: this result is just the value 
of the magnetic field computed from the quantum fluctuations over the protogalactic 
scale.

\renewcommand{\theequation}{4.\arabic{equation}}
\setcounter{equation}{0}
\section{Conducting initial conditions}
\label{sec4}
The most generic protoinflationary initial conditions are characterized by a globally neutral plasma consisting of an equal number of positively and negatively charged species whose total energy density and pressure are given by:
\begin{equation}
\overline{\rho}_{\mathrm{tot}} = \overline{\rho}_{+} + \overline{\rho}_{-} + \overline{\rho}_{r}, \qquad 
\overline{p}_{\mathrm{tot}} = \overline{p}_{+} + \overline{p}_{-} + \overline{p}_{r}.
\label{cond1}
\end{equation}
As in the previous section,  the bar over a given quantity indicates that the corresponding variable is physical (as opposed to comoving).  
In what follows $m_{\pm}$ and $\overline{T}_{\pm}$ shall denote the masses and the temperatures 
of either the positive or negative charge carriers. As long as $ \overline{T}_{\pm} \gg m_{\pm}$, the temperatures 
$\overline{T}_{+}$ and $\overline{T}_{-}$ coincide with $\overline{T}_{r}$ which is, by definition, the temperature of the radiation, i.e. $\overline{T}_{+} \simeq \overline{T}_{-} \simeq \overline{T}_{r}$. In the opposite case (i.e. for $\overline{T}_{\pm} < m_{\pm}$) the evolution of the various temperatures 
depends on the plasma parameter $g_{\mathrm{plasma}}$ coinciding with the inverse of the number of charge carriers contained in the Debye sphere.  
\subsection{Thermodynamical considerations}
The different pressures and energy densities appearing in Eq. (\ref{cond1}) are:
\begin{eqnarray}
&& \overline{\rho}_{\pm} = m_{\pm} \overline{n}_{\pm} + \frac{3}{2}\, \overline{n}_{\pm} \, \overline{T}_{\pm}, \qquad 
\overline{p}_{\pm} = \overline{n}_{\pm} \overline{T}_{\pm},
\label{cond2}\\
&& \overline{\rho}_{r} = \frac{\pi^2}{15} \, g_{\mathrm{th}}\, \overline{T}_{r}^4, \qquad \overline{p}_{r} = \frac{\pi^2}{45} \,  g_{\mathrm{th}}\, \overline{T}_{r}^4.
\label{cond3}
\end{eqnarray}
From Eqs. (\ref{cond1}), (\ref{cond2}) and (\ref{cond3}) the first principle of the thermodynamics and the adiabaticity of the evolution imply the following relation:
\begin{equation} 
d \biggl[ V_{H} \,( \overline{n}_{+} m_{+} + \overline{n}_{-} m_{-}) + \frac{3}{2} \biggl( \overline{n}_{+} \overline{T}_{+} + 
\overline{n}_{-} \overline{T}_{-} \biggr) + a^3 \overline{\rho}_{r} \biggr] + \biggl( \overline{n}_{+} \overline{T}_{+} + 
\overline{n}_{-} \overline{T}_{-} + \overline{p}_{r} \biggr) d V_{H} =0,
\label{cond4}
\end{equation}
where $V_{H} = (4 \pi/3) H_{*}^{-3} a^3$ and $H_{*}^{-3}$ is a fiducial 
volume coinciding, for instance, with the volume of the event horizon at the onset of the inflationary stage.  
Assuming the global neutrality of the plasma, the concentrations of the charged species must be equal, i.e.  
$\overline{n}_{+} = \overline{n}_{-} = \overline{n}_{0}$ and Eq. (\ref{cond4}) becomes:
\begin{equation}
d[ a^2 ( \overline{T}_{+} + \overline{T}_{-})] + a\, \gamma\, d (a \overline{T}_{r} ) =0, \qquad \gamma = \frac{2 s}{n_{0}},
\label{cond5}
\end{equation}
where $n_{0} = a^3 \, \overline{n}_{0} $ and $s =a^3 \, \overline{s}$ are, respectively, the comoving concentration of the charged species and the comoving entropy density. Equation 
(\ref{cond5}) can be solved under different approximations. The physical initial conditions stipulate that, 
 initially, $\overline{T}_{+} \simeq \overline{T}_{-} \simeq 
\overline{T}_{r}$ with the result that the common temperature of the different species scales as\footnote{Recall that 
$\zeta(3) =  1.202$.}
\begin{equation}
\overline{T} \simeq a^{- \frac{4 + \gamma}{2 + \gamma}},\qquad \gamma = \frac{4 \pi^4}{45\, \zeta(3)} \, \biggl(\frac{n_{r}}{n_{0}}\biggr),
\label{cond6}
\end{equation}
where $n_{r} = a^3 \overline{n}_{r}$ and $\overline{n}_{r} = g_{\mathrm{th}} \, \overline{T}_{r}^3 \,\zeta(3)/\pi^2$.
If $n_{r} \ll n_{0}$, $\overline{T}$ scales, approximately, as $a^{-2}$ in the opposite case (i.e. $n_{r} \ll n_{0}$) the effective temperature evolves, to first order in $1/\gamma$, as $a^{-1}$.

\subsection{Plasma approximation and breaking of duality symmetry}
At finite charge density the gauge-invariant action of the system is
\begin{equation}
S_{\mathrm{tot}} = S_{Y} -\int \sqrt{- g} \, j^{(+)}_{\mu} Y^{\mu}\, d^4 x +  \int \sqrt{- g} \, j^{(-)}_{\mu} Y^{\mu}\, d^4 x + S_{\mathrm{charged}} + S_{\mathrm{neutral}},
\label{eq1}
\end{equation}
where $S_{Y}$ has been given in Eq. (\ref{act1}) and $j_{\mu}^{(\pm)}$ denote four-currents 
associated with the charge carriers; $S_{\mathrm{charged}} = S_{(+)} + S_{(-)}$ 
and $S_{\mathrm{neutral}}$ denote, respectively, the actions of the charged and neutral species.  
By minimizing the action (\ref{eq1}) the evolution equations can be obtained and then presented in terms 
of the normal modes introduced in Eq. (\ref{Tmn3}):
\begin{eqnarray}
&& \frac{1}{\sqrt{\lambda}} \vec{\nabla} \cdot ( \sqrt{\lambda} \, \vec{E}) = 4 \pi q (n_{+} - n_{-}),\qquad 
\sqrt{\lambda} \vec{\nabla} \cdot \biggl( \frac{\vec{B}}{\sqrt{\lambda}}\biggr) =0,
\label{eq2a}\\
&& \frac{1}{\sqrt{\lambda}} \vec{\nabla} \times (\sqrt{\lambda} \, \vec{B} ) = 4 \,\pi \,q ( n_{+} \vec{v}_{+} - n_{-} \vec{v}_{-}) + \frac{1}{\sqrt{\lambda}} \frac{\partial}{\partial \tau}( \sqrt{\lambda} \, \vec{E}), 
\label{eq3}\\
&& \sqrt{\lambda} \vec{\nabla} \times \biggl(\frac{\vec{E}}{\sqrt{\lambda}} \biggr) = 
- \sqrt{\lambda}  \frac{\partial}{\partial \tau} \biggl(\frac{\vec{B}}{\sqrt{\lambda}}\biggr),
\label{eq4}
\end{eqnarray}
where\footnote{According to the rescaling established in the 
previous section, we should have that $q \to q/(4\pi)$ in Eqs. (\ref{eq1}) and (\ref{eq3}). We  prefer, however, to redefine $e$ and keep the Gaussian units where the $4\pi$ appear throughout since these conventions are the standard ones for the 
analysis of weakly coupled plasmas.} $q= e/\sqrt{\lambda}$. The normal modes 
of Eq. (\ref{Tmn3}) have been used in section \ref{sec3} and in appendix 
\ref{APPA} in the source-free case.  It is then natural to employ the same variables also at finite 
charge density. 

The source terms in Eqs. (\ref{eq2a}) and (\ref{eq3}) break the duality symmetry discussed in Eqs. (\ref{I7}) and (\ref{I8}). The main difference between quantum and classical  initial conditions resides in the breaking of the duality symmetry. In the case of the quantum initial conditions the duality symmetry is broken at the end of the 
inflationary phase. Conversely for classical (or conducting) initial conditions the duality symmetry is broken already during the protoinflationary phase. Even if Eqs. (\ref{eq2a}) and (\ref{eq3})--(\ref{eq4}) hold  in general, for direct comparison with the results of section \ref{sec3}
the attention will be focussed on the case of homogeneous $\sqrt{\lambda}$.
 
 Assuming an approximate kinetic equilibrium between the charged species, the first of Eq. (\ref{eq2a}) implies
$- \nabla^2 \phi = 4 \pi q ( n_{+} - n_{-})$ with $n_{\pm}(\phi) = n_{0} \exp{[ \mp q \phi/T_{\pm}]}$ where $\phi$ denotes the Coulomb potential; expanding $n_{\pm}(\phi)$ for $|q \phi/T_{\pm}|<1$,  we get $\nabla^2 \phi - \ell_{\mathrm{D}}^{-2} \phi =0$. The explicit form of the comoving Debye screening length $\ell_{\mathrm{D}}$ is 
\begin{equation}
 \ell_{\mathrm{D}} = \sqrt{\frac{T_{+}\, T_{-}}{4\pi q^2 n_{0}(T_{+} + T_{-})}}\to \sqrt{\frac{T}{8\pi e^2 n_{0}}}\, \sqrt{\lambda}.
\label{eq8}
\end{equation}
The second expression of Eq. (\ref{eq8}) defining $\ell_{\mathrm{D}}$ holds in the limit of approximate
thermal equilibrium between the charged species, i.e. $T_{+} \simeq T_{-} \to T$.
In the presence of a background of charged species the correct expansion parameter 
is not the gauge coupling itself but rather the inverse of the total number of particles present inside the Debye sphere:
\begin{equation}
g_{\mathrm{plasma}} = \frac{1}{N_{D}}, \qquad N_{D} = \frac{4}{3} \,\pi\, n_{0}\, \ell_{D}^3,\qquad \ell_{\mathrm{D}} = \sqrt{\frac{T}{8\, \pi q^2 \, n_{0}}};
\label{cond7}
\end{equation}
$g_{\mathrm{plasma}}$ plays the role of coupling constant of the plasma and it measures the degree to which collective effects dominate over single particle behaviour. In other words the plasma parameter is small when many particles interact at the same time.  For the Debye shielding to occur the number of particles in a Debye sphere must be large. But this means $g_{\mathrm{plasma}} \ll 1$ and the latter condition defines the conventional plasma approximation. 

The validity of the plasma approximation ensures that the system 
can have a charge density, a shielded electric field but very limited interaction 
of single particles. It is easy to check that $\gamma$ appearing in Eq. (\ref{cond6}) can be written in terms of $1/g_{\mathrm{plasma}}^2$ showing, as anticipated, that the corrections to the adiabatic scaling 
 of the effective temperature, proportional to $1/\gamma$, are indeed 
 small in the plasma approximation. If Weyl invariance is unbroken the plasma parameter is also Weyl invariant. 
Since the plasma parameter does not explicitly depend on the masses of the charge carriers  the following chain of proportionality relations holds: 
\begin{equation}
g_{\mathrm{plasma}} \propto (n_{0} \, \ell_{\mathrm{D}}^3)^{-1}
\propto (n_{0} \, T^{3})^{-1/2};
\end{equation}
but $n_{0} \, T^{3}$ is insensitive to the expansion as it is 
evident by passing from comoving to physical quantities: $ (n_{0} \, T^{3})^{-1/2} =  (\overline{n}_{0} \, \overline{T}^{3})^{-1/2}$. The Weyl invariance 
of the plasma parameter implies that the initial value of 
 $g_{\mathrm{plasma}}$ will be preserved by the dynamical evolution.
 
If Weyl invariance is broken the validity of the perturbative expansion and the 
increase of the Debye shielding scale can be used as criteria for 
enforcing the consistency of the perturbative approximation and the 
consequent screening of all electric fields throughout quasi-de Sitter 
stage.  According to Eqs. (\ref{eq8}) and (\ref{cond7}) $\ell_{\mathrm{D}}$ must increase with $\sqrt{\lambda}$ since 
only in this case the electric fields will be progressively screened 
as the plasma evolves during the protoinflationary phase. But this is indeed 
the same requirement necessary for the decrease of $g_{\mathrm{plasma}}$.
In the opposite case (i.e. when $\sqrt{\lambda}$ decreases) the electric fields 
are not properly screened so that the Debye scale at the end of inflation is actually smaller than the value it had at the onset of inflation.  

The validity of the plasma approximation implies a preferred evolution for the Debye shielding length. It is therefore possible to trade the evolution 
of $\sqrt{\lambda}$ for the evolution of $\ell_{\mathrm{D}}$. On a 
phenomenological ground, it seems reasonable to adopt, as working guess, the following expression\footnote{ Note that $a_{r}$ denotes the conventional normalization of the scale factor at the reheating epoch. The evolution 
 of the Debye shielding increases during the protoinflationary phase and also during the inflationary phase. The 
 intermediate exponential suppression at reheating may be absent but has been included to account for a possible breaking of the plasma approximation in the transition regime.} 
\begin{equation}
\lim_{a_{i} \leq a\ll a_{\mathrm{r}}} \frac{\ell_{\mathrm{D}}(a)}{\ell_{\mathrm{D}}(a_{i})} = \sqrt{\frac{\lambda}{\lambda_{i}}}
\to \biggl(\frac{a}{a_{i}}\biggr)^{\nu -1/2},\qquad \lim_{a \sim a_{\mathrm{r}}} \frac{\ell_{\mathrm{D}}(a)}{\ell_{\mathrm{D}}(a_{i})} = \sqrt{\frac{\lambda}{\lambda_{i}}} \to e^{- a/a_{r}};
\label{LL1}
\end{equation}
for $a \gg a_{\mathrm{r}}$,  $\ell_{\mathrm{D}}\to \mathrm{constant}$, 
the plasma is again Weyl invariant (at least as long as the charge carriers are relativistic) but duality is still broken at finite conductivity. In the case of quasi-de Sitter dynamics, Eq. (\ref{LL1}) reproduces, by construction,  the evolution of $\sqrt{\lambda}$ already discussed in section \ref{sec3} in the complementary case of quantum and thermal initial conditions.  
This coincidence facilitates the comparison between the two sets of initial conditions.
\subsection{Lorentzian plasmas and conductivity}
Even if $\sqrt{\lambda}$ does not evolve in time, the Weyl symmetry is broken because of the mass of the charge carriers. In section \ref{sec2}  it has been shown that Ohmic currents are Weyl invariant in the relativistic limit 
where masses and chemical potentials are negligible in comparison with the temperature. 
The contribution of the charged species to the transport coefficients can be computed  within a kinetic model of the plasma. Working under the hypothesis that the collisions between the particles of the same charge can be neglected, it is consistent  to assume that $\Gamma_{\pm} \gg \Gamma_{+}$ and $\Gamma_{\pm} \gg \Gamma_{-}$, as it happens for Lorentzian plasmas \cite{KT,KTA}; furthermore, at high temperatures $\Gamma_{\pm} > H$  for interactions mediated by massless gauge bosons. 

The Vlasov-Landau equation for the positively and negatively charged species is given by:
\begin{equation}
  \frac{\partial f_{\pm}}{\partial \tau}  + \vec{v}_{\pm} \cdot \vec{\nabla}_{\vec{x}} f_{\pm} +\frac{d \vec{P}_{\pm}}{d\tau}\cdot
  \vec{\nabla}_{\vec{p}} f_{\pm} =\biggl( \frac{\partial f_{\pm}}{\partial \tau }\biggr)_{\mathrm{coll}},
  \label{VL1}
\end{equation}
where $\vec{v}_{\pm}$ and $\vec{p}_{\pm}$ are, respectively, the comoving three-velocities and the comoving three-momenta defined as: 
\begin{equation}
 \frac{d \vec{P}_{\pm}}{d\tau} = \pm q [ \vec{E} + \vec{v}_{\pm} \times \vec{B}], \qquad \vec{v}_{\pm} = \frac{\vec{P}_{\pm}}{\sqrt{P_{\pm}^2 + m_{\pm}^2 \, a^2 }}.
\label{VL3}
\end{equation}
If $P_{\mathrm{\pm}}^2 \gg m^2 a^2$ the whole system is invariant under Weyl rescaling and the evolution 
of the geometry does not affect directly the evolution of the charged species. In the opposite limit the two relations 
appearing in Eq. (\ref{VL3})  can be combined and the equations for the velocity field become:
\begin{eqnarray}
&& \vec{v}_{+}^{\prime} + {\mathcal H} \,\vec{v}_{+} = 
\frac{q \, n_{+}}{\rho_{+} a} [ \vec{E} + \vec{v}_{+} \times 
\vec{B}] + a \Gamma_{\pm}  \frac{\rho_{+}}{\rho_{-}} (\vec{v}_{-} - \vec{v}_{+}) + 
\frac{4}{3} \frac{\rho_{r}}{ \rho_{+}} a \Gamma_{+\,r} (\vec{v}_{r} - \vec{v}_{+}),
\label{VL4}\\
&& \vec{v}_{-}^{\prime} + {\mathcal H} \,\vec{v}_{-} = 
-\frac{q \overline{n}_{-}}{\overline{\rho}_{-} a} [ \vec{E} + \vec{v}_{-} \times 
\vec{B}] +a \Gamma_{\mp} \frac{\rho_{-}}{\rho_{+}} (\vec{v}_{+} - \vec{v}_{-}) +  
\frac{4}{3} \frac{\rho_{r}}{ \rho_{-}} 
a \Gamma_{-\, r} (\vec{v}_{r} - \vec{v}_{-}),
\label{VL5}
\end{eqnarray}
where $\vec{v}_{r}$ is the velocity of the radiation fluid whose dynamics is not relevant for the present ends; $\Gamma_{\pm\, r}$ are the rates of momentum exchange between charged and neutral species while $\Gamma_{\pm}$ are the rates of momentum exchange among the charged species themselves. 

To deduce the expression of the conductivity  Eqs. (\ref{VL4}) and (\ref{VL5}) 
must be combined to obtain the evolution of the center of mass velocity and the equation for the total current:
\begin{equation}
\vec{v} = \frac{m_{+} \, \vec{v}_{+} + m_{-} \, \vec{v}_{-}}{m_{+} + m_{-}}, \qquad \vec{J} = q n_{0} (\vec{v}_{+}
- \vec{v}_{-}).
\end{equation}
The result of the mentioned combination is:
\begin{eqnarray}
&& \vec{v}^{\,\prime} + {\mathcal H}  \vec{v} = \frac{\vec{J}\times 
\vec{B}}{a^4 \overline{\rho}(1 + m_{-}/m_{+})}+ \frac{4}{3}
 \frac{\overline{\rho}_{r}}{\overline{\rho}} a \Gamma_{r\, -} (\vec{v}_{r} - \vec{v}),
\label{VL6}\\
&& \vec{J}^{\prime} + \biggl({\mathcal H} + a\Gamma_{\mathrm{\pm}} + 
\frac{4 \rho_{\gamma}\Gamma_{- r}}{ 3 n_{0} \,m_{\mathrm{e}}}\biggr) \vec{J} = 
\frac{\omega_{\mathrm{p}}^2}{4\pi} \biggl(\vec{E} + \vec{v} \times \vec{B} + \frac{\vec{\nabla} p_{-}}{q\,n_{0}} - 
\frac{\vec{J}\times\vec{B}}{q n_{0}}\biggr)
\nonumber\\
&& + \frac{4 q \rho_{r} \Gamma_{r -}}{3 m_{-}} (\vec{v} - \vec{v}_{r}).
 \label{VL7}
\end{eqnarray}
where $\rho = \rho_{+} + \rho_{-}$ and $\omega_{\mathrm{p}}$ is the plasma frequency.  To simplify the expressions a hierarchy in the masses of the  charge carriers can be assumed (for instance $m_{+} > m_{-} = m$). In Eq. (\ref{VL7}) the terms $\vec{J}'$ and ${\mathcal H} \vec{J}$ are 
comparable in magnitude and are both smaller than $\Gamma_{\pm}$ and $\Gamma_{\pm r}$. Thus 
the explicit form of the Ohm law becomes:
\begin{equation}
\vec{J} = \sigma \biggl(\vec{E} + \vec{v} \times \vec{B} + \frac{\vec{\nabla} p_{-}}{q\,n_{0}}- \frac{\vec{J}\times \vec{B}}{n_{0} q}\biggr),\qquad \sigma = \frac{\omega_{\mathrm{p}}^2 }{4\pi a \Gamma_{\mathrm{\pm}} },
\label{VL8}
\end{equation}
where the contribution of the neutral species has been neglected\footnote{In what follows the Hall term and the thermoelectric term will be considered to be of higher order and they will be anyway irrelevant for 
the present ends.}. For a Lorentzian plasma \cite{KT,KTA}, the conductivity can 
then be written as:
\begin{equation}
\sigma(a,\gamma) = \frac{T(a,\gamma)}{q^2 \sqrt{1 + \frac{m a}{T(a,\gamma)}}},\qquad \lim_{\gamma \gg 1} T(a,\gamma) \propto a^{- 2/(\gamma +2)} = \mathrm{constant}.
\label{ss0}
\end{equation}
Consider first the case when the comoving Debye length does not evolve 
(i.e. $\ell_{\mathrm{D}}$ constant or, in equivalent terms, $\nu = 1/2$ in Eq. (\ref{LL1})).
In the limit $\gamma \gg 1$ the comoving temperature $ T = \overline{T} a$ is approximately constant and this is verified provided $g_{\mathrm{plasma}} \ll 1$.  In the limit $T \gg m a$ Eq. (\ref{ss0}) implies $\sigma \simeq T/q^2$ (as it happens in the case of a relativistic plasma); in the opposite limit, $\sigma \simeq T/q^2 \sqrt{T/(m a)}$. 

When $\ell_{\mathrm{D}}$ evolves as in Eq. (\ref{LL1}), the breaking of the Weyl invariance produced by the finite 
mass of the charge carriers interferes with the evolution of the Debye shielding. After expressing Eq. (\ref{ss0})  in terms of the number of inflationary efolds $N = \ln{(a/a_{\mathrm{*}})}$, the evolution of the conductivity
becomes:
\begin{equation}
\sigma(N) = \sigma_{*}\,\frac{e^{(N + N_{i}) (2 \nu -1)}}{\sqrt{1 + e^{N - N_{\mathrm{c}}}}};
\label{sigma1}
\end{equation}
note that $N_{i}$ accounts for the initial number of efolds during the protoinflationary phase while $N_{\mathrm{c}}$ is given by:
\begin{equation}
N_{\mathrm{c}} = -0.253+ \frac{1}{2} \ln{\xi} - \frac{1}{4} \ln{g_{\mathrm{th}}} - \ln{\biggl(\frac{m}{M_{\mathrm{P}}}\biggr)}.
\label{sigma2}
\end{equation}
Following the notations 
of Eq. (\ref{LL1}), Eq. (\ref{sigma1}) holds before reheating, i.e. in the regime where initial conditions are set. 
To leading order in $g_{\mathrm{plasma}}$ the critical number of efolds $N_{\mathrm{c}}$  depends on the temperature reached during the protoinflationary phase and it can be estimated by recalling that at the onset of the inflationary phase the total energy density does not exceed the contribution of the protoinflationary 
energy density stored, for instance, in relativistic species.  With this logic,  from Eq. (\ref{sigma2}), 
an upper bound on $N_{\mathrm{c}}$ can be derived and it can be expressed as
\begin{equation}
N_{\mathrm{c}}  \simeq 36.78 - 0.25 \ln{(g_{\mathrm{th}}/100)} + 0.5 \ln{(\xi/10^{-5})} - \ln{(m/\mathrm{GeV})}.
\label{sigma3}
\end{equation}
Given the value of $N_{\mathrm{c}}$ there are four physically distinct situations.
If  $N \simeq N_{\mathrm{tot}} \gg N_{\mathrm{c}} + N_{\mathrm{min}}$ and $\nu=1/2$, Weyl invariance is broken before the onset of the last ${\mathcal O}(63)$ efolds of inflationary expansion; this means, in practice, that the sources do not contribute to the initial conditions which are accurately fixed by quantum mechanics. 
If $N_{\mathrm{min}}< N \leq N_{\mathrm{c}} + N_{\mathrm{min}}$ and $\nu=1/2$ the conductivity will be constant for the first $N_{\mathrm{c}}$ efolds and then it will be exponentially suppressed as 
$e^{(N_{\mathrm{c}} -N_{\mathrm{min}})/2}$ (if $N\sim N_{\mathrm{\mathrm{min}}}$) and as $e^{-N_{\mathrm{min}}/2}$ (if $N\sim N_{\mathrm{min}} + N_{\mathrm{c}}$). If $\nu \simeq 3/4$ the conductivity is practically constant for $N\gg N_{\mathrm{c}}$;
Finally if $\nu \neq 3/4$ and $\nu \neq 1/2$ the evolution depends on the specific value of $\nu$.

The goal here is not to endorse (or predict) a specific duration of the inflationary phase but just to convey the message  
that when $N_{\mathrm{min}}< N \leq N_{\mathrm{c}} + N_{\mathrm{min}}$ the last ${\mathcal O}(63)$ efolds of inflationary expansion may start when the conductivity did not undergo a substantial  suppression. For instance, when the mass range of the lightest charge carrier is ${\mathcal O}(\mathrm{GeV})$ and if $N\sim {\mathcal O}(N_{\mathrm{min}})$ the conductivity is still almost constant ${\mathcal O}(30)$ efolds prior to the end of inflation. In this class of physical situations the normalization of the electric and magnetic fields does not follow from the quantum mechanical initial conditions but rather from the conducting initial conditions.  
\subsection{Plasma initial conditions}
 Equations (\ref{ss0}) and 
(\ref{sigma1}) imply that for $N < N_{\mathrm{c}}$ the conductivity 
can be assumed to be roughly constant only if $\nu = 1/2$. When $\nu \neq 1/2$ Eqs. (\ref{eq2a}), (\ref{eq3}) and (\ref{eq4}) must be numerically integrated to determine the evolution of the electric and magnetic fields. Using the normalized scale factor as evolution parameter Eqs. (\ref{eq2a}), (\ref{eq3}) and (\ref{eq4}) can be reduced to the 
following pair of vector equations\footnote{Equations (\ref{INC1}) and (\ref{INC2}) are supplemented 
by the requirement that $\vec{\nabla}\cdot\vec{E} = \vec{\nabla}\cdot \vec{B} =0$ as implied in the case 
of a globally neutral electric plasma.}
\begin{eqnarray}
\frac{\partial \vec{E}}{\partial \alpha} &=& - \frac{{\mathcal F}}{\alpha {\mathcal H}} \vec{E}  - \frac{\vec{J}}{\alpha {\mathcal H}} + \frac{\vec{\nabla} \times \vec{B}}{\alpha {\mathcal H}},
\label{INC1}\\
\frac{\partial \vec{B}}{\partial \alpha} &=& \frac{{\mathcal F}}{\alpha {\mathcal H}} \vec{B} - \frac{\vec{\nabla} \times \vec{E}}{\alpha {\mathcal H}},
\label{INC2}
\end{eqnarray}
where $\alpha = a/a_{*}$; $a_{*}$ denotes conventionally the value of the scale factor when the inflationary phase starts and ${\mathcal H}$ is given by:
\begin{equation}
{\mathcal H} = {\mathcal H}_{*} \sqrt{\biggl(\frac{a}{a_{*}}\biggr)^{3 w +1} + \biggl(\frac{a_{*}}{a}\biggr)^2}.
\label{INC3}
\end{equation} 
According  to Eq. (\ref{INC3}) the protoinflationary phase is dominated by a perfect fluid with barotropic index $w$; the inflationary phase is simply modeled by an effective cosmological constant. 
More complicated models of protoinflationary evolution can be formulated by introducing scalar field sources
together with perfect fluids. A class of analytic solutions of the Friedmann equations with this property is reported in appendix \ref{APPC}.
\begin{figure}[!ht]
\centering
\includegraphics[height=7cm]{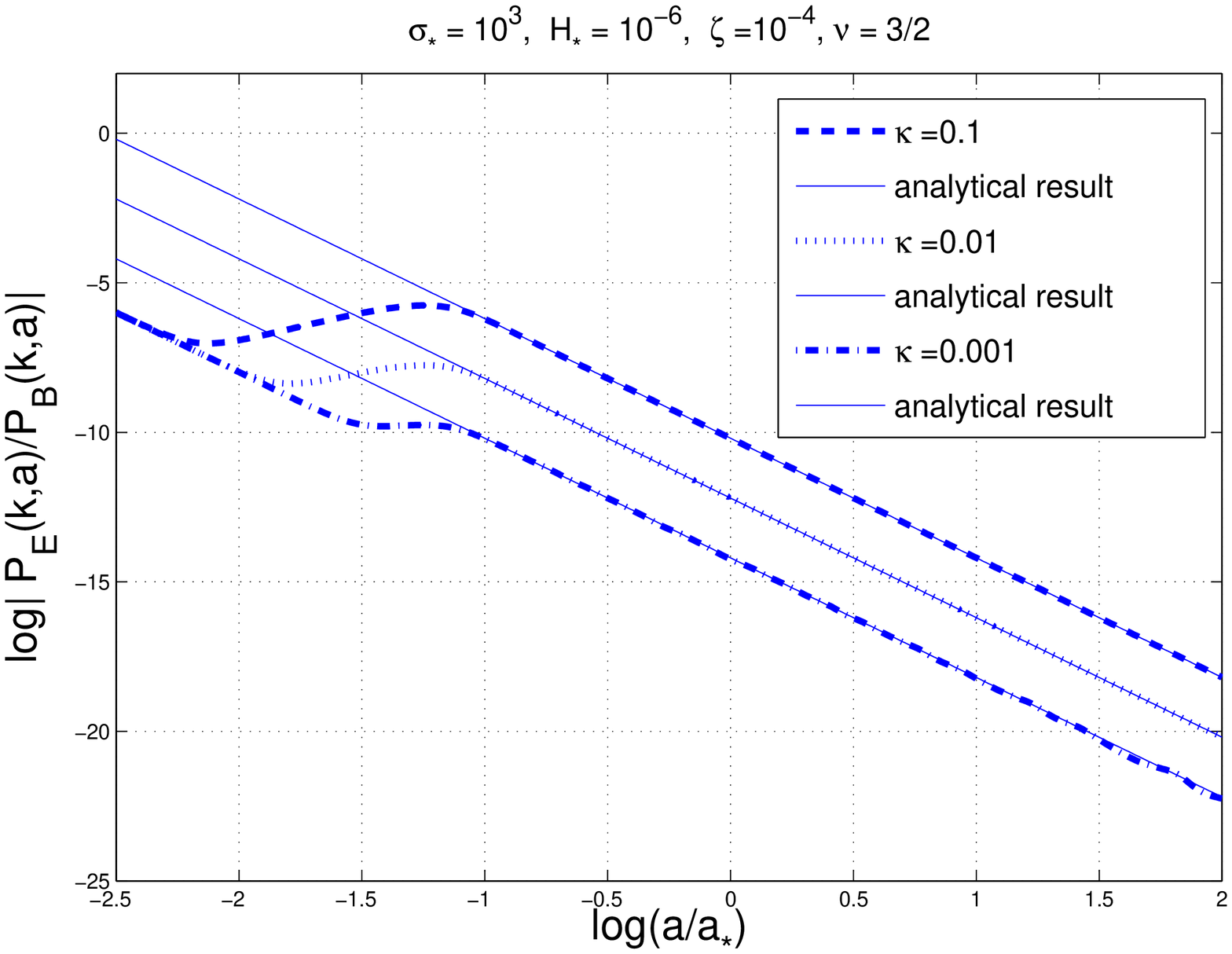}
\caption[a]{The result of the numerical integration for the case $w=1/3$.}
\label{figureS1}      
\end{figure}
For the present purposes, the details of the inflationary sources are immaterial. 
This has been explicitly checked by using the solution illustrated in Eqs. (\ref{SOL1})
and (\ref{SOL2}). 

\begin{figure}[!ht]
\centering
\includegraphics[height=7cm]{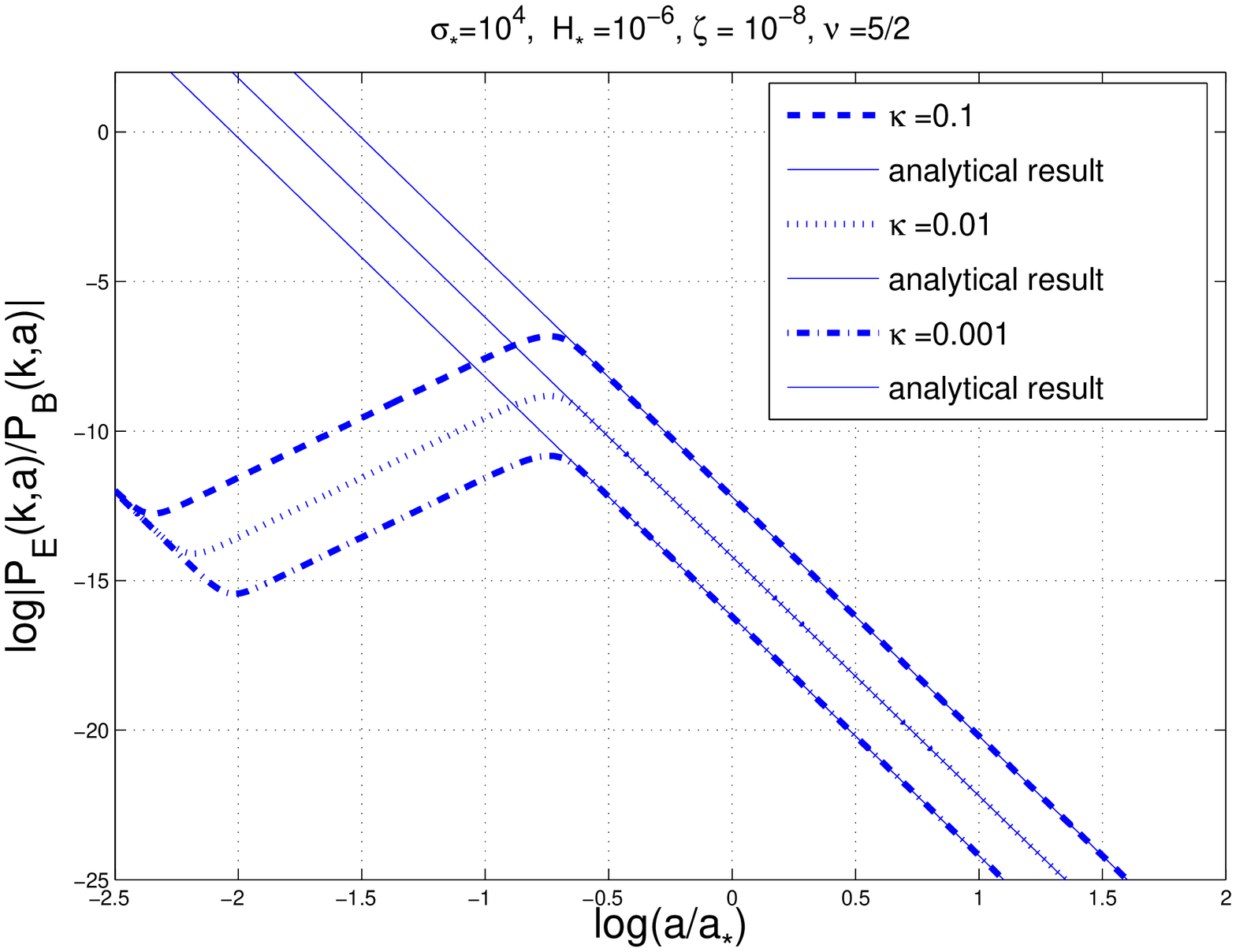}
\caption[a]{The result of the numerical integration for the case $w=1$.}
\label{figureS2}      
\end{figure}

Following the previous discussions and, in particular, Eq. (\ref{LL1}) 
 Eqs. (\ref{INC1}) and (\ref{INC2}) can be integrated 
conductivity evolves as
\begin{equation}
\sigma(\alpha) = \frac{\sigma_{*} (a/a_{i})^{ 2 \nu -1}}{\sqrt{1 + (a/a_{*})\zeta}}, \qquad \zeta = e^{- N_{\mathrm{c}}};
\label{SOL7}
\end{equation}
$\zeta$ depends on the specific value of $N_{\mathrm{c}}$. By arbitrarily assuming 
that $\sigma$ is constant for $N< N_{\mathrm{c}}$ we can expect that
the ratio between the electric and the magnetic power spectra approximately 
obeys the scaling law:
\begin{equation}
\lim_{\alpha \gg 1} \frac{P_{\mathrm{E}}(k,\alpha)}{P_{\mathrm{B}}(k,\alpha)} \to \frac{k^2}{ 16 \pi^2 \sigma^2(a)},
\label{SOL8}
\end{equation}
valid in the limit  $a \gg a_{*}$. The numerical results confirm this guess.
In Figs. \ref{figureS1} and \ref{figureS2} the system (\ref{INC1}) and (\ref{INC2}) has been integrated for two illustrative 
values of the barotropic index, i.e. $w=1/3$ and $w=1$. On the vertical axis, in both figures, the common logarithm of the ratio between the electric and the magnetic power spectra is reported; on the horizontal axis we have the common logarithm of the normalized scale factor.  The value of $H_{*}$ is given in Planck 
units while $\sigma_{*}$ is essentially the normalization of the conductivity in units of $H_{*}$. In both figures the dashed, dotted and dot-dashed lines correspond to three different values of $\kappa = k/H_{*}$ which is
the wavenumber in units of $H_{*}$. The full (thin) lines correspond to the analytical approximation 
of Eq. (\ref{SOL8}) valid for $a\gg a_{*}$ but plotted also at earlier times just to guide the eye. 

In the limit where all the fields appearing in Eqs. (\ref{INC1}) and (\ref{INC2}) are solenoidal (i.e.  $\vec{\nabla}\cdot\vec{E} = \vec{\nabla} \cdot \vec{B} = \vec{\nabla}\cdot \vec{J} =0$) the displacement current can be neglected for very high conductivity and therefore the appropriate initial conditions for the electromagnetic fields at $\tau_{x}$ are simply given by 
\begin{equation}
\vec{B}(\vec{x},\tau_{x}) = \vec{B}^{(\mathrm{in})}(\vec{x}), \qquad \vec{E}(\vec{x},\tau_{x}) = \frac{\vec{\nabla} \times
\vec{B}^{(\mathrm{in})}(\vec{x})}{4 \pi \sigma(a_{x})},
\label{sol2a}
\end{equation}
as assumed, without explicit numerical proof, in \cite{weyl}. The power spectra  subjected to the conducting initial 
conditions (\ref{sol2a}) and subsequently amplifield according to Eqs. (\ref{LL1}) and (\ref{INC1})--(\ref{INC2}) are\footnote{The dimensionless variables $x$ and $z$ introduced hereunder and appearing in Eqs. (\ref{solNN10}), (\ref{solNN11}) and (\ref{sol12}) must not be confused with the dimensionless variables of section \ref{sec3}. Since the two sets 
of variables never appear simultaneously, potential confusions are avoided.}:
\begin{equation}
P_{\mathrm{B}}(k,\tau,\tau_{x}) = P_{\mathrm{B}}(k,\tau_{x}) \, |{\mathcal U}(z, x)|^2, \qquad 
P_{\mathrm{E}}(k,\tau,\tau_{x}) = P_{\mathrm{B}}(k,\tau_{x}) \, |{\mathcal V}(z, x)|^2
\label{solNN10}
\end{equation}
where $x = k \tau_{x}$,  $z= - k \tau$ and $\eta = k/4\pi \sigma$; the expressions  
\begin{eqnarray}
{\mathcal V}(x,z) &=& \frac{i \pi}{4} \, \sqrt{\frac{x}{z}} \biggl\{ \frac{1}{x}\biggl[ P_{\nu}^{(1)}(z) P_{\nu}^{(2)}(x) 
- P_{\nu}^{(1)}(x) P_{\nu}^{(2)}(z)\biggr]  
\nonumber\\
&-& \eta \biggl[ H_{\nu}^{(2)}(x) P_{\nu}^{(1)}(z) - H_{\nu}^{(1)}(x) P_{\nu}^{(2)}(z)\biggr]\biggr\},
\nonumber\\
{\mathcal U}(x,z) &=& \frac{i \pi}{4} \, \sqrt{\frac{z}{x}}\biggl\{ \biggl( 2 \nu - \eta \, x \biggr) \biggl[H_{\nu}^{(2)}(x) 
H_{\nu}^{(1)}(z) - H_{\nu}^{(1)}(x) H_{\nu}^{(2)}(z) \biggr]
\nonumber\\
&+& x \biggl[ H_{\nu + 1}^{(1)}(x) H_{\nu}^{(2)}(z) - H_{\nu+1}^{(2)}(x) H_{\nu}^{(1)}(z)\biggr]\biggr\},
\label{solNN11}
\end{eqnarray}
are given in terms of the Hankel functions of first and second kind (i.e. 
$H_{\nu}^{(1)}(z)$ and $H_{\nu}^{(2)}(z)$) as well as
in terms of the following combinations:
\begin{equation} 
 P_{\nu}^{(1)}(z) = 2 \nu H_{\nu}^{(1)}(z) - z H_{\nu + 1}^{(1)}(z), \qquad P_{\nu}^{(2)}(z) = 2 \nu H_{\nu}^{(2)}(z) - z H_{\nu + 1}^{(2)}(z).
 \label{sol12}
 \end{equation}
By setting $\tau= - \tau_{x}$ the expressions of Eq. (\ref{solNN10}) reproduce the conducting initial conditions given in Eq. (\ref{SOL8}).  
Expanding ${\mathcal V}(x, z)$ and ${\mathcal U}(x, z)$ 
in the limit $x \ll 1$ the following expressions can be obtained:
\begin{eqnarray}
{\mathcal V}(x, z) &=& x^{-\nu} \biggl[ \sqrt{z} {\mathcal V}_{1}(x, z) + {\mathcal O}\biggl(x^{7/2}\biggr)\biggr]
+ x^{\nu}  \biggl[ \sqrt{z} {\mathcal V}_{2}(x, z) + {\mathcal O}\biggl(x^{9/2}\biggr)\biggr],
\label{SOLINT1}\\
{\mathcal U}(x, z) &=& x^{-\nu} \biggl[ \sqrt{z} {\mathcal U}_{1}(x, z) + {\mathcal O}\biggl(x^{7/2}\biggr)\biggr]
+ x^{\nu}  \biggl[ \sqrt{z} {\mathcal U}_{2}(x, z) + {\mathcal O}\biggl(x^{9/2}\biggr)\biggr]
\label{SOLINT2}
\end{eqnarray}
where the following functions have been introduced:
\begin{eqnarray}
{\mathcal V}_{1}(x, z) &=& 2^{\nu - 3}\, \sqrt{x} \, J_{\nu}(z) \biggl[ x ( \eta x - 2) \Gamma(\nu -1) + 4 \eta \Gamma(\nu)\biggr],
\nonumber\\
{\mathcal V}_{2}(x, z) &=&\frac{2^{- 5 -\nu} \, \pi \, J_{- \nu}(z)[ 32\nu (\nu +1) - 16
\eta ( \nu + 1) x - 8 (\nu +1) x^2 + 4 \eta x^3 + x^4]}{\sqrt{x} \Gamma(\nu + 2) \sin{(\pi \nu)}}
\nonumber\\
{\mathcal U}_{1}(x, z) &=& 2^{\nu -3}\, \sqrt{x} \, J_{\nu-1}(z) [ 4 \eta \Gamma(\nu) + \Gamma(\nu -1) x ( \eta x - 2)],
\nonumber\\
{\mathcal U}_{2}(x, z) &=&\frac{2^{- 5 -\nu} \, \pi \, J_{1- \nu}(z)[8\Gamma(\nu +2) ( ( x + 2 \eta) x - 4 \nu ) - \Gamma(\nu +1) x^3 ( x + 4 \eta)]}{\sqrt{x} \Gamma(\nu +1) \Gamma(\nu + 2) \sin{(\pi \nu)}},
\label{SOLINT3}
\end{eqnarray}
where $J_{\nu}(z)$ is the ordinary Bessel function.
For $|\tau_{x}| \ll |\tau_{e}|$ (and $x <1$, $z<1$) the electric power spectra are suppressed throughout 
the whole stage of inflationary expansion. 
Note that the rate of the suppression is larger than in the quantum and thermal cases examined in section \ref{sec3}.
From Eqs. (\ref{SOLINT1}), (\ref{SOLINT2}) and (\ref{SOLINT3}), at the end of the 
inflationary phase, $P_{\mathrm{B}}(k,\tau_{e},\tau_{x}) \simeq P_{\mathrm{B}}(k,\tau_{x}) (a_{e}/a_{x})^{2\nu -1}$
and $P_{\mathrm{E}}(k,\tau_{e},\tau_{x}) \simeq \eta P_{\mathrm{B}}(k,\tau_{x}) \, (a_{e}/a_{x})^{1 - 2\nu}$.
Since $\eta = k/(4\pi \sigma) \ll 1$ the suppression 
of the electric fields is always much larger than in the case of vacuum initial conditions.
\begin{figure}[!ht]
\centering
\includegraphics[height=7cm]{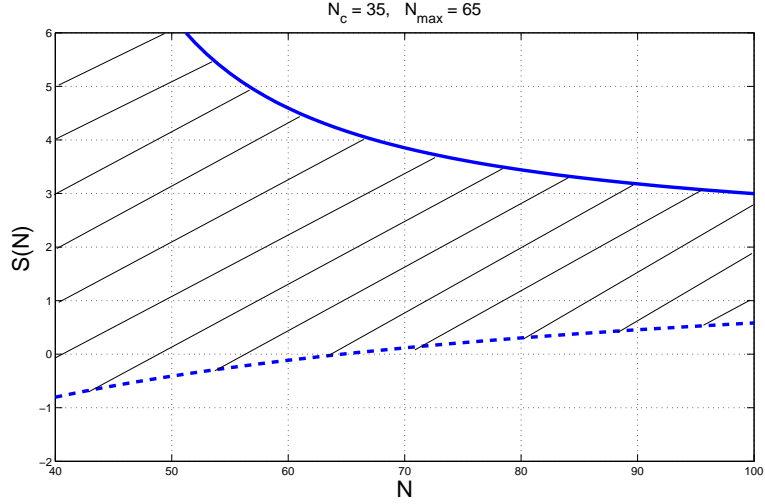}
\caption[a]{The bounds of Eq. (\ref{BOUND}) are illustrated for $N_{\mathrm{c}} = 35$ and $N_{\mathrm{max}} =65$.}
\label{figureS3}      
\end{figure}

The amplitude of the magnetic power spectrum during the protoinflationary phase can be maximized by only asking the compatibility with the closure bounds. In the latter case the power spectrum of the magnetic field will be given by 
\begin{equation}
\frac{\sqrt{P_{{\mathcal B}}(k,\tau_{x}, \tau_{0})}}{\mathrm{Gauss}} = 10^{-60.15} \biggl(\frac{\Xi}{10^{-4}}\biggr)^{1/2} \biggl(\frac{H_{r}}{H_{e}}\biggr)^{2 \alpha_{1} - 1}  e^{ (\nu -1/2) (N - N_{\mathrm{c}}) - 2 (N - N_{\mathrm{max}})},
\label{sol15}
\end{equation}
where $\Xi=  \sqrt{P_{{\mathcal B}}(k,\tau_{1})}/(H_{1}^2 M_{\mathrm{P}}^2) < 1$ measures the fraction of energy density stored in the magnetic field at $\tau_{1}$; $H_{r}$ accounts for the possibility of a delayed radiation-dominated phase between the end of inflation and the onset of big-bang nucleosynthesis. The exponent $\alpha_{1}$ depends on the expansion rate between the end 
of the inflationary phase and the onset of the standard (i.e. post-inflationary) radiation-dominated epoch. Equation (\ref{sol15}) has several interesting limits.
In the case 
$\nu = 1/2$ there is no amplification due to the evolution of the gauge coupling and therefore the upper bound on the 
magnetic field intensity is around ${\mathcal O}(10^{-61})$ Gauss in the sudden reheating approximation where $H_{r} \sim H_{e}$. It is interesting that this 
figure coincides with what has been obtained in Eq. (\ref{sol13}) for $\nu= 1/2$,
$k =0.1 \, \mathrm{Mpc}^{-1}$ and $N \gg N_{\mathrm{max}}$.
This result coincides with the magnetic field one would obtain from the 
protoinflationary initial conditions in the case of standard thermal history and 
minimal number of efolds.  If $\nu = 5/2$ and $N = N_{1} \simeq {\mathcal O}(65)$ and $N_{\mathrm{c}} \sim {\mathcal O}(35)$  the maximal magnetic field 
turns out to be  $10^{-35}$ Gauss (in the sudden reheating approximation) which can become of the order of ${\mathcal O}(10^{-23})$ Gauss for a stiff post-inflationary phase extending down to the nucleosynthesis scale.

\begin{figure}[!ht]
\centering
\includegraphics[height=7cm]{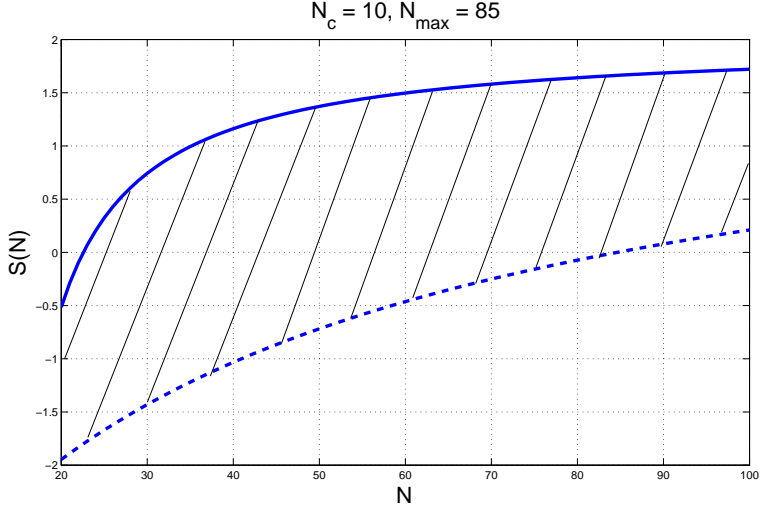}
\caption[a]{The bounds of Eq. (\ref{BOUND}) are illustrated for $N_{\mathrm{c}} = 10$ and $N_{\mathrm{max}} =85$.}
\label{figureS4}      
\end{figure}

The classical and the quantum results can be more quantitatively compared by imposing, for instance, the following reasonable hierarchy of inequalities:
\begin{equation}
\biggl(\frac{\sqrt{P_{{\mathcal B}}(k,\tau_{0})}}{\mathrm{Gauss}}\biggr)_{\mathrm{quantum}} \leq 
\biggl(\frac{\sqrt{P_{{\mathcal B}}(k,\tau_{0})}}{\mathrm{Gauss}}\biggr)_{\mathrm{classical}} \leq 
\biggl(\frac{\sqrt{P_{{\mathcal B}}(k,\tau_{0})}}{\mathrm{Gauss}}\biggr)_{\mathrm{maximal}}.
\label{BOUND}
\end{equation}
The quantum and classical contributions both have to be smaller than some maximal value which is computed by assuming that the magnetic power spectrum is of the order of the energy density of the inflaton at the end of inflation. The results of this comparison are reported in Figs. \ref{figureS3} and \ref{figureS4}
for different values of $N_{\mathrm{c}}$ and $N_{\mathrm{max}}$. On the vertical axis 
the value of the normalized rate is reported and a function of the total number of efolds, i.e. $S(N) = {\mathcal F}/{\mathcal H}$ which coincides with $(\nu -1/2)$ during the quasi-de Sitter stage of expansion. With the full line the bound stemming from the maximal value of the magnetic power spectrum is indicated. 
For values of $S(N)$ and $N$ within the shaded region the hierarchy expressed by Eq. (\ref{BOUND}) is satisfied. Below the dashed line, in both plots, the first inequality of Eq. (\ref{BOUND}) is inverted and the quantum contribution is larger than the classical one. Figures \ref{figureS3} and \ref{figureS4} are 
purely illustrative and they show that both classical and quantum initial conditions lead to sizable 
magnetic fields. The differences between the two sets of initial conditions must come from a more 
thorough analysis of the spectral properties especially for typical length-scales 
larger than the Mpc. 
\renewcommand{\theequation}{5.\arabic{equation}}
\setcounter{equation}{0}
\section{Monopole plasma}
\label{sec5}
The duality symmetry can be used 
to deduce what happens if instead of an electric current we have a monopole current. 
In a globally neutral monopole plasma the value of the total magnetic charge vanishes i.e. $n_{+} = n_{-}$ where now $n_{+}$ and $n_{-}$ denote the monopole concentrations with opposite charges.
The masses of the positively and negatively charged monopoles are assumed to be different (i.e. $m_{+} \neq m_{-}$). There is no possible confusion with the notations used for the electric plasma since we are not going to consider here dyons but simply a globally neutral plasma of magnetic monopoles. In the past the possibility of a dyonic plasma in magnetohydrodynamics has been discussed 
in \cite{olesmagn1,olesmagn2}; the situation considered here is, by construction, 
even simpler since only magnetic charges will be considered.
Even if this is strictly not necessary, the monopoles are assumed to be non-relativistic both during the protoinflationary phase and during the inflationary phase. The transport coefficients can be computed by exploiting the assumption of Lorentzian plasma \cite{KT} in full analogy with what has been discussed in section \ref{sec4} 
but with the crucial difference that the role of the magnetic and of the electric degrees of freedom is somehow 
interchanged. The relevant system of equations is given by:
\begin{eqnarray}
&& \nabla_{\mu} ( \lambda \, Z^{\mu\nu}) = 0, \qquad \nabla_{\mu} \tilde{Z}^{\mu\nu} = 4 \pi \ell^{\nu}, 
\label{mon2}\\
&& m_{\pm} \biggl[ \frac{d u_{\pm}^{\mu}}{ds} + \Gamma^{\mu}_{\alpha\beta} \,u^{\alpha}_{\pm} \,u^{\beta}_{\pm} \biggr] = \pm \tilde{e} \tilde{Z}^{\mu}_{nu} \, u^{\nu}_{\pm}, 
\label{mon3}
\end{eqnarray}
where $\tilde{e}$ denotes  the monopole charge.  The explicit form of Eq. (\ref{mon2}) 
becomes then:
\begin{eqnarray}
&& \frac{1}{\sqrt{\lambda}} \vec{\nabla}\cdot ( \sqrt{\lambda} \vec{E} ) =0,\qquad \sqrt{\lambda} \vec{\nabla} \cdot ( \frac{\vec{B}}{\sqrt{\lambda}}) = 4 \pi \tilde{q} \,(n_{+} - n_{-}),
\label{mon1a}\\
&& \frac{1}{\sqrt{\lambda}} \vec{\nabla} \times (\sqrt{\lambda} \, \vec{B} ) = \frac{1}{\sqrt{\lambda}} \frac{\partial}{\partial \tau}( \sqrt{\lambda} \, \vec{E}),
\label{mon1c}\\
&& \sqrt{\lambda} \vec{\nabla} \times \biggl(\frac{\vec{E}}{\sqrt{\lambda}} \biggr) + \sqrt{\lambda}  \frac{\partial}{\partial \tau} \biggl(\frac{\vec{B}}{\sqrt{\lambda}}\biggr) = - 4 \pi  \tilde{q} (n_{+} \vec{v}_{+} - n_{-} \vec{v}_{-}),
\label{mon1d}
\end{eqnarray}
where $\tilde{q} = \frac{\tilde{e}}{\sqrt{\lambda}}$; note that $\vec{v}_{\pm}$ denote the velocities of monopoles with opposite charges. From Eq. (\ref{mon3}) the evolution of the comoving three-momenta can be derived and it is given by
\begin{equation}
\frac{d \vec{P}_{\pm}}{d \tau} = \pm \tilde{q} \biggl[ \vec{B} - \vec{v}_{\pm} \times \vec{E}\biggr], \qquad \vec{v}_{\pm} = \frac{\vec{P}_{\pm}}{\sqrt{P_{\pm}^2 + m^2 a^2}}.
\label{mon5}
\end{equation}
In the relativistic limit $P^2 \gg m^2 a^2$, $\vec{v}_{\pm} = \vec{P}_{\pm}/|\vec{P}|$ . In the case $P^2 \ll m^2 a^2$ the evolution equations of the velocity will be manifestly not invariant under Weyl rescaling:
\begin{eqnarray}
&& \vec{v}_{+}' + {\mathcal H} \vec{v}_{+} = 
\frac{\tilde{q} \,\overline{n}_{+}}{\overline{\rho}_{+} a} [ \vec{B} - \vec{v}_{+} \times 
\vec{E}] + a \tilde{\Gamma}_{\pm}  \frac{\rho_{+}}{\rho_{-}} (\vec{v}_{-} - \vec{v}_{+}) + 
\frac{4}{3} \frac{\overline{\rho}_{r}}{ \overline{\rho}_{+}} a \overline{\Gamma}_{+\,r} (\vec{v}_{r} - \vec{v}_{+}),
\label{mon5a}\\
&& \vec{v}_{-}' + {\mathcal H} \vec{v}_{-} = 
-\frac{\tilde{q}\, \overline{n}_{-}}{\overline{\rho}_{-} a} [ \vec{B} - \vec{v}_{-} \times 
\vec{E}] +a \tilde{\Gamma}_{\mp} \frac{\rho_{-}}{\rho_{+}} (\vec{v}_{+} - \vec{v}_{-}) +  
\frac{4}{3} \frac{\overline{\rho}_{r}}{ \overline{\rho}_{-}} 
a \tilde{\Gamma}_{-\, r} (\vec{v}_{r} - \vec{v}_{-}).
\label{mon6}
\end{eqnarray}
 Defining the center of mass 
velocity of the monopole system and neglecting the momentum exchange between the monopoles 
and the radiation background the following pair of equations can be obtained:
\begin{equation}
 \vec{v}_{\mathrm{m}}' + {\mathcal H} \vec{v}_{\mathrm{m}} = - \frac{\vec{J}_{\mathrm{m}} \times \vec{E}}{a^4 \overline{\rho}_{\mathrm{m}}}, \qquad \vec{J}_{\mathrm{m}} = \sigma_{\mathrm{m}}( \vec{B} - \vec{v}_{\mathrm{m}} \times \vec{E}),
\label{monv2}
\end{equation}
where $\vec{J}_{\mathrm{m}} = n_{\mathrm{m}} (\vec{v}_{+} - \vec{v}_{-})$ is the total monopole 
current and $\sigma_{\mathrm{m}}$ is the magnetic conductivity. 
The main equations of the electric and magnetic fields in the presence of the monopole current are given by
\begin{eqnarray}
&& \sqrt{\lambda} \partial_{\tau} \biggl[ \frac{1}{\lambda} \partial_{\tau} \biggl( 
\sqrt{\lambda} \vec{E} \biggr) \biggr] - \nabla^2 \vec{E} = - 4 \pi \vec{\nabla}\times 
\vec{J}_{\mathrm{m}}, 
\label{CC}\\
&& \frac{1}{\sqrt{\lambda}} \partial_{\tau} \biggl[ \lambda \partial_{\tau} \biggl( \frac{\vec{B}}{\sqrt{\lambda}} \biggr) \biggr] - \nabla^2 \vec{B} = - \frac{4 \pi}{\sqrt{\lambda}} \partial_{\tau}\biggl(\sqrt{\lambda} \vec{J}_{\mathrm{m}} \biggr).
\label{DD}
\end{eqnarray}
We can assume, as before, that $\sqrt{\lambda}$ is homogeneous.
The situation is dual to the case of the electric charges: the magnetic fields are screened while the electric flux is conserved at finite magnetic conductivity. The induced magnetic field will then be $\vec{B} \simeq -\vec{\nabla}\times \vec{E}/(4 \pi \sigma_{\mathrm{m}})$. It is interesting to speculate 
that an initial monopole plasma could screen the magnetic components but not the 
electric ones which could be instead amplified if $\sqrt{\lambda}$ decreases (to keep the magnetic field screened); the electric components may be converted back to magnetic fields at reheating. The analysis of these themes is beyond the aims of this section whose only purpose is to illustrate the role of the duality symmetry in connection with the initial conditions of inflationary magnetogenesis.
\renewcommand{\theequation}{6.\arabic{equation}}
\setcounter{equation}{0}
\section{Concluding remarks}
\label{sec6}
Which are the correct initial conditions for protoinflationary magnetogenesis? Are they classical? Are they thermal?
Are they quantum mechanical? Are plasma sources immaterial when setting the initial conditions 
of inflationary magnetogenesis? Are the obtainable magnetic fields phenomenologically relevant? 
These are some of the main questions addressed in the present investigation whose motivation stems from the observation
that vacuum initial conditions for the evolution of large-scale gauge fields are customarily imposed 
regardless of the number of inflationary efolds and in spite of the possible presence of protoinflationary remnants. It will be recorded that conventional inflation is known to be geodesically incomplete in the past and while it is true that the contribution 
of the sources to the evolution of the geometry is likely to be exponentially suppressed with the number of efolds,
the same conclusion does not apply to the evolution of large-scale gauge fields. 

The symmetries of inflationary magnetogenesis suggest
that the most generic initial conditions are neither quantum nor thermal  but rather conducting: if charged particles are not fine-tuned to vanish, the simplest situation compatible with the phenomenological constraints is 
to contemplate the case of a globally neutral protoinflationary plasma. 
Conventional inflation takes place when the gravitational coupling is strong. As inflation proceeds, the temperature of the protoinflationary plasma decreases at a rate which depends on the smallness of the plasma parameter reflecting the largeness of the Debye shielding scale which must increase if magnetic fields are to be amplified and electric fields are screened. In this situation the electric fields are suppressed in comparison with the magnetic fields not only asymptotically in the future but also 
at the onset of the protoinflationary phase. 

The Weyl and the duality symmetries determine the initial conditions of inflationary magnetogensis since, as usual, the symmetries of the problem are reflected in the symmetries of the solutions.
Quantum initial conditions do not break explicitly the duality symmetry but may break the Weyl symmetry if the gauge kinetic term is coupled to a spectator field. Classical initial conditions may not break the Weyl symmetry but break explicitly the duality symmetry since they imply, in the simplest case, the presence of Ohmic currents without a corresponding magnetic source.
The present considerations demonstrate explicitly that both conducting and quantum mechanical initial conditions 
lead to phenomenologically relevant large-scale magnetic fields so that possible distinctions between the two sets of initial conditions must rely on subtle differences in the spectral properties for typical length-scales much larger than the one of the protogalactic collapse. 
\newpage
\begin{appendix}
\renewcommand{\theequation}{A.\arabic{equation}}
\setcounter{equation}{0}
\section{Duality-invariant Hamiltonian}
\label{APPA}
The duality symmetry leaves the Hamiltonian invariant but when the Hamiltonian depends explicitly on time (as in the present case) every canonical transformation changes its form. 
Which is the correct Hamiltonian to use for the quantization and for the evolution 
of the electromagnetic fluctuations?  The punchline of the following appendix is that there indeed exist different Hamiltonians, all differing by a canonical transformations,  but only one class of Hamiltonians is explicitly invariant under duality and this will be the preferred one, for the present purposes. 

In the Coulomb gauge\footnote{The Coulomb gauge condition is preserved under 
Weyl rescaling of the time-dependent metric; the Lorentz 
gauge condition does not have the same property; see the discussion after Eq. (\ref{WW4b}).}, and in flat space-time, 
the duality properties have been discussed in \cite{duality1,duality2} where it has been 
noted that even if it is possible to keep both $Y_{0}$ and the longitudinal part of the Abelian vector potential (i.e. $\vec{Y}_{\mathrm{L}}$), the transverse 
variables are decoupled from the longitudinal  ones and from the gauge contributions. Without any gauge fixing, the extremization of the action with respect to $Y_{0}$ implies that $\vec{Y}^{\,\,'}_{\mathrm{L}} = \vec{\nabla} Y_{0}$ 
(see \cite{duality1,duality2}). Thus the action of Eq. (\ref{act1})  can then be written, in the Coulomb gauge, as
\begin{equation}
S_{Y} = \int \,d\tau\, L_{Y}(\tau), \qquad L_{Y}(\tau) = \int d^{3} x\, {\mathcal L}_{Y}(\vec{x},\tau),
\label{dual5}
\end{equation}
with
\begin{equation}
{\mathcal L}_{Y}(\vec{x},\tau) = \frac{1}{2} \biggl\{ \vec{y}^{\,\prime \,2} + {\mathcal F}^2 
 \vec{y}^{\,2}  - 2 {\mathcal F} \vec{y} \cdot \vec{y}^{\,\prime} - \partial_{i} \vec{y} \cdot \partial^{i} \vec{y}\biggr\}.
\label{dual6}
\end{equation}
The canonical momentum conjugate to $\vec{y}$ can be obtained from Eq. (\ref{dual6}) and it coincides, up to a sign, with the canonical electric field defined in Eq. (\ref{Tmn3}), i.e. $\vec{\pi} = \vec{y}^{\,\prime} - {\mathcal F} \vec{y} = - \vec{E}$.
The canonical Hamiltonian is simply obtained from the Lagrangian density 
$H_{Y}(\tau) = \int d^{3} x\, [\vec{\pi} \cdot \vec{y} - {\mathcal L}_{Y}(\vec{x},\tau)]$ and its explicit 
form is: 
\begin{equation}
H_{Y}(\tau) = \frac{1}{2} \int d^3 x \biggl[ \vec{\pi}^{2} + 2 \,{\mathcal F} \, \vec{\pi} \cdot \vec{y} + 
\partial_{i} \vec{y} \cdot \partial^{i} \vec{y}\biggr].
\label{dual8}
\end{equation}
The Fourier mode expansion for the canonical fields
\begin{equation}
\vec{\pi}(\vec{x},\tau) = \frac{1}{(2\pi)^{3/2}} \int d^{3} k\,\, \vec{\pi}_{\vec{k}}(\tau) \,\,e^{-i \vec{k}\cdot\vec{x}}, \qquad 
 \vec{y}(\vec{x},\tau) = \frac{1}{(2\pi)^{3/2}} \int d^{3} k \,\, \vec{y}_{\vec{k}}(\tau) \,\,e^{-i \vec{k}\cdot\vec{x}},
 \label{dual9}
 \end{equation}
can be inserted into Eq. (\ref{dual8}) and the resulting expression is:
\begin{equation}
H_{Y}(\tau) = \frac{1}{2} \int d^3 k \biggl[ \vec{\pi}_{\vec{k}} \cdot \vec{\pi}_{-\vec{k}} +  {\mathcal F}
\biggl( \vec{\pi}_{\vec{k}} \cdot \vec{y}_{-\vec{k}}  +  \vec{\pi}_{-\vec{k}} \cdot \vec{y}_{\vec{k}}\biggr)
+k^2 \vec{y}_{\vec{k}} \cdot \vec{y}_{-\vec{k}}\biggr].
\label{dual10}
\end{equation}
The Hamiltonian (\ref{dual10}) is invariant under the transformation $\sqrt{\lambda} \to 1/\sqrt{\lambda}$ provided, at the 
same time the electric variables are appropriately rotated into the magnetic ones and vice-versa, i.e.
\begin{eqnarray}
 \vec{\pi}_{\vec{k}} \to - k \,\vec{y}_{\vec{k}}, \qquad \vec{y}_{\vec{k}} \to \frac{1}{k} \,\vec{\pi}_{\vec{k}},\qquad  \vec{\pi}_{-\vec{k}} \to - k \,\vec{y}_{-\vec{k}}, \qquad \vec{y}_{-\vec{k}} \to \frac{1}{k}\, \vec{\pi}_{-\vec{k}},
\label{dual11}
\end{eqnarray}
where $k = |\vec{k}|$. The Lagrangian 
given in Eq. (\ref{dual6}) is not invariant under the duality transformation of Eq. (\ref{dual11}) and this 
happens since the duality transformation maps a tensor into a pseudo-tensor. The Hamilton equations derived from Eq. (\ref{dual10}) become:
\begin{equation}
\vec{y}_{\vec{k}}^{\,\prime} = \vec{\pi}_{\vec{k}} + {\mathcal F} \,\vec{y}_{\vec{k}},\qquad
\vec{\pi}_{\vec{k}}^{\,\prime} = - k^2 \,\vec{y}_{\vec{k}} - {\mathcal F}\, \vec{\pi}_{\vec{k}}.
\label{dual13}
\end{equation}
Under the transformation of Eq. (\ref{dual11}) the two equations of Eq. (\ref{dual13}) are transformed one into the other 
and vice-versa. Since the Hamiltonian (\ref{dual10}) is time-dependent its form can be changed by a canonical transformation. The Hamiltonian (\ref{dual10}) is invariant under the transformation (\ref{dual11}) exactly because the canonical momenta are, up to a sign, the canonical electric fields.  If the Hamiltonian is different, the explicit duality invariance is not guaranteed. Consider, indeed, the following functional of the old fields (i.e. $\vec{y}$), of the new momenta (i.e. $\vec{\Pi}$) and of the conformal time:
\begin{equation}
{\mathcal R}[\vec{y}, \vec{\Pi}, \tau] = \int d^3\, x\biggl[ \vec{y}\cdot \vec{\Pi} - \frac{{\mathcal F}}{2} \vec{y}\cdot\vec{y}\biggr],
\label{GEN1}
\end{equation}
The old momenta are related to the new ones as $\vec{\pi} = \vec{\Pi} - {\mathcal F} \vec{y}$ while 
the new Hamiltonian becomes $H_{Y}(\tau) \to H^{(\mathrm{new})}_{Y}(\tau) = H_{Y}(\tau) + \partial_{\tau} {\mathcal R}$, i.e. 
\begin{equation}
H^{(\mathrm{new})}_{Y}(\tau) = \frac{1}{2} \int d^3 x \biggl[ \vec{\Pi}^2 - \biggl( {\mathcal F}^2 + {\mathcal F}'\biggr) \vec{y}^2 
+ \partial_{i} \vec{y} \cdot \partial^{i} \vec{y} \biggr].
\label{GEN3}
\end{equation}
If the calculations are carried on in terms of $H^{(\mathrm{new})}_{Y}(\tau)$ (and not, as preferable, in terms of 
$H_{Y}(\tau)$)  the spectra of $\vec{\Pi}$ will not be related by duality transformations to the spectra 
of $\vec{B} = \vec{\nabla}\times \vec{y}$ simply because $\vec{\Pi}$ does not coincide with the canonical electric 
field $\vec{E}$. It is therefore mathematically convenient and physically justified to use $H_{Y}(\tau)$ rather than $H^{(\mathrm{new})}_{Y}(\tau)$. In the literature this point
is never mentioned but it seems essential when computing the magnetic and the electric power spectra with either quantum or thermal normalization.

\renewcommand{\theequation}{B.\arabic{equation}}
\setcounter{equation}{0}
\section{Mode functions: explicit expressions}
\label{APPB}
The solution for the mode functions can be discussed with different techniques 
depending on the monotonicity properties of $\lambda$.
If $\lambda$ has a monotonic dependence on the conformal time coordinate, 
the evolution equations for $f_{k}(\tau)$ and $g_{k}(\tau)$ can be solved 
exactly. Suppose, for instance, that $\sqrt{\lambda} = \sqrt{\lambda}_{1} ( - \tau/\tau_{1})^{1/2 - \nu}$. Thus, the rate of variation of $\sqrt{\lambda}$ is given by ${\mathcal F} =(1/2 - \nu)/\tau$ and 
Eq. (\ref{DD21}) can be solved in terms of Hankel functions:
\begin{eqnarray}
&& f_{k}(\tau) = \frac{{\mathcal N}}{\sqrt{2 k}} \, \sqrt{- k \tau} \, H_{\nu}^{(1)}(- k \tau),\qquad {\mathcal N} = \sqrt{\frac{\pi}{2}} e^{i \pi( \nu +1/2)/2},
\label{Solf1}\\
&& g_{k}(\tau) = - {\mathcal N}\, \sqrt{\frac{k}{2}}\, \sqrt{-k\tau} \, H_{\nu-1}^{(1)}(- k \tau). 
\label{Solg1}
\end{eqnarray}
The dual of $f_{k}$, $g_{k}$ and ${\mathcal F}$ are, respectively,  
$ f_{k} \to g_{k}/k$, $g_{k} \to  - k f_{k}$ and  ${\mathcal F} \to  - {\mathcal F}$.
Duality acts non-trivially on the solutions  
given in Eqs. (\ref{Solf1}) and (\ref{Solg1}). Indeed, if ${\mathcal F} \to \tilde{{\mathcal F}} = (1/2 - \mu)/\tau$, 
$\tilde{{\mathcal F}} = - {\mathcal F}$ provided $ \nu = 1 - \mu$. Duality can then be used, ultimately, to relate the spectra 
of the electric and magnetic fields in the ideal (and somehow unphysical) 
situation where the finite conductivity effects are absent both 
at the end and at the beginning of inflation. 

Whenever the evolution of $\sqrt{\lambda}$ is not monotonic or not solvable analytically in a closed form, Eq. (\ref{DD21}) can be separately solved in the limits 
$k/{\mathcal F} > 1$ and $k/{\mathcal F} <1$. When $k/{\mathcal F} > 1$ 
\begin{equation}
f_{k}(\tau) = \frac{1}{\sqrt{2 k}} e^{- i k (\tau - \tau_{\mathrm{in}})},\qquad g_{k}(\tau) = - i \sqrt{\frac{k}{2}} e^{- i k (\tau - \tau_{\mathrm{in}})},
\label{app1}
\end{equation}
valid for $\tau \le \tau_{\mathrm{ex}}$ where, by definition, $\tau_{\mathrm{ex}}$ is  ${\mathcal F}(\tau_{\mathrm{ex}}) = k$.
In the limit $k/{\mathcal F} < 1$ the solution of Eqs. (\ref{DD21})  can be formally expanded in powers of the spatial gradients:
\begin{equation}
f_{k}(\tau) = \sum_{\ell =0}^{\infty} k^{2 \ell} f_{k,\, \ell}(\tau), \qquad 
g_{k}(\tau) = \sum_{\ell =0}^{\infty} k^{2 \ell} g_{k,\, \ell}(\tau),
\label{app2}
\end{equation}
with the result that $f_{k,\,\ell}$ and $g_{k,\,\ell}$ obey the following hierarchy of coupled equations:
\begin{eqnarray}
&& f_{k,\,\ell}' = g_{k,\, \ell} + {\mathcal F} \,f_{k,\, \ell}, \qquad \ell \geq 0,
\label{app3}\\
&& g_{k,\,0}' = - {\mathcal F}\, g_{k,\, 0} , \qquad \ell =0,
\label{app4}\\
&& g_{k,\,\ell}' = - f_{k,\, \ell -1} - {\mathcal F} \,g_{k,\, \ell}, \qquad \ell \geq 1.
\label{app5}
\end{eqnarray}
Equations (\ref{app3})--(\ref{app5}) are analytically solvable order by order and for a generic form 
of the pump field. The lowest order solution with the boundary conditions 
of Eq. (\ref{app1}) is simply 
\begin{eqnarray}
&& f_{k}(\tau) = \frac{1}{\sqrt{2 k}} \biggl[ \sqrt{\frac{\lambda(\tau)}{\lambda_{\mathrm{ex}}}} - 
i k \sqrt{\lambda_{\mathrm{ex}}\, \lambda(\tau)}\, {\mathcal I}(\tau_{\mathrm{ex}},\tau) \biggr] \, e^{- i k (\tau_{\mathrm{ex}} - 
\tau_{\mathrm{in}})}, 
\label{app6}\\
&& g_{k}(\tau) = - i \sqrt{\frac{k}{2}} \,\sqrt{\frac{\lambda_{\mathrm{ex}}}{\lambda(\tau)}} \, e^{- i k (\tau_{\mathrm{ex}} - 
\tau_{\mathrm{in}})},\qquad {\mathcal I}(\tau_{\mathrm{ex}},\tau) = \int_{\tau_{\mathrm{ex}}}^{\tau} \, \frac{d\tau'}{\lambda(\tau')},
\label{app7}
\end{eqnarray}
where $\lambda_{\mathrm{ex}} = \lambda(\tau_{\mathrm{ex}})$. 
When the Universe reheats, the conductivity of the plasma breaks the duality symmetry and the relevant equations are given by:
\begin{equation}
g_{k}' = - k^2 f_{k} - 4 \pi \sigma \,g_{k} , \qquad f_{k}' = g_{k}.
\label{Soll4}
\end{equation}
Defining with $\overline{f}_{k}$ and $\overline{g}_{k}$ the solutions of Eqs. (\ref{DD21}) 
the solutions for $\tau \geq \tau_{\sigma}$ can be obtained rather easily by direct matching with the result that 
\begin{eqnarray}
f_{k}(\tau) &=& \frac{e^{- z (\tau,\tau_{\sigma})}}{k\, \alpha(k,\sigma)} \biggl\{ k \overline{f}_{k}(\tau_{\sigma})  \biggl[ \alpha(k,\sigma) 
\cosh{[y(\tau,\tau_{\sigma})]} + \Sigma(k,\sigma) \sinh{[y(\tau,\tau_{\sigma})]} \biggr]
\nonumber\\
&+& \overline{g}_{k}(\tau_{\sigma}) \sinh{[y(\tau,\tau_{\sigma})]}  \biggr\},
\label{app8}\\
g_{k}(\tau) &=& \frac{e^{- z (\tau,\tau_{\sigma})}}{\alpha(k,\sigma)} \biggl\{  \overline{g}_{k}(\tau_{\sigma}) 
 \biggl[ \alpha(k,\sigma) 
\cosh{[y(\tau,\tau_{\sigma})]} - \Sigma(k,\sigma) \sinh{[y(\tau,\tau_{\sigma})]} \biggr] 
\nonumber\\
&-& k \overline{f}_{k}(\tau_{\sigma}) \sinh{[y(\tau,\tau_{\sigma})]}\biggr\},
\label{app9}
\end{eqnarray}
where 
\begin{equation}
\alpha(k,\tau) = \sqrt{\Sigma^2(\tau)-1}, \qquad \Sigma(\tau) = \frac{2\pi\,\sigma(\tau)}{k}.
\label{app9a}
\end{equation}
Note that $\Sigma(\tau)$  measures the ratio of the conductivity over the wavenumber. 
In Eqs. (\ref{app8}) and (\ref{app9}) $y(\tau,\tau_{\sigma})$ and $z(\tau,\tau_{\sigma})$ depend on the 
evolution of the conductivity and are defined in Eq. (\ref{app10}). From Eqs. (\ref{app8}) and (\ref{app9})
$|f_{k}(\tau)|^2$ and $|g_{k}(\tau)|^2$ are easily obtained 
and they can be studied in the limit $k\ll 2 \pi \sigma$ (i.e. $\Sigma \gg 1$) 
where the conductivity dominates. The opposite case (i.e. when $k\gg 2 \pi \sigma$) is
obtainable from Eqs. (\ref{app8}) and (\ref{app9}) by appreciating that, in this limit, 
$\alpha \to i \beta$, $y \to i \tilde{y}$ and $z \to i \tilde{z}$ where $\beta = \sqrt{1 - \Sigma^2}$. In the latter situation, the expressions of $|f_{k}(\tau)|^2$ and $|g_{k}(\tau)|^2$ are different since $\alpha$, $y$ and $z$ become all 
complex quantities.  It is finally appropriate to remark that the evolution of the canonical operators 
is such that $[\hat{y}_{i}, \hat{\pi}_{j}] \to 0$ for $\tau \geq \tau_{\sigma}$.
Because of the presence of the conductivity the Wronskian is driven to zero. This 
means that out of the two solutions of the system only one survives, i.e. the one 
related to the magnetic part.  The vanishing of the Wronskian signals 
the transition to the classical dynamics where the magnetic field operators become 
Gaussian random fields.

\renewcommand{\theequation}{C.\arabic{equation}}
\setcounter{equation}{0}
\section{Protoinflationary evolution: analytic example}
\label{APPC}
An explicit solution describing the protoinflationary dynamics is \cite{mxpr}:
\begin{eqnarray}
&& a(t) = a_{*} \biggl[ \sinh{(\beta \, H_{*}\, t)}\biggr]^{1/\beta}, \qquad \beta = \frac{3 ( w + 1)}{2},
\label{SOL1}\\
&& \varphi(t) = \varphi_{0} \pm \sqrt{\frac{2}{\beta}} \overline{M}_{\mathrm{P}} \sqrt{1 - \Omega_{*}} \ln{\biggl[\tanh{\biggl( \frac{\beta H_{*} t}{2}\biggr)}\biggr]}, 
\label{SOL2}
\end{eqnarray}
where $w$ is the barotropic index characterizing the protoinflationary fluid while $a(t)$ and $\varphi(t)$ are the scale factor 
and the inflaton expressed in cosmic time;
the parameter $\Omega_{*}$ measures the fluid fraction of the protoinflationary energy density 
\begin{equation}
\Omega_{*}= \frac{\rho_{*}}{3 H_{*}^2 \overline{M}_{\mathrm{P}}^2},\qquad \overline{\rho}_{\mathrm{tot}}(t)= \rho_{*} \biggl(\frac{a_{*}}{a}\biggr)^{ 3 ( w +1)},
\label{SOL4a}
\end{equation}
where $\rho_{\mathrm{tot}}$ denotes the protoinflationary energy density.
The potential for $\varphi$ is given by
\begin{equation} 
V(\varphi) = 3 H_{*}^2 \overline{M}_{\mathrm{P}}^2 + 
\frac{3}{2} (1 - w) H_{*}^2 \overline{M}_{\mathrm{P}}^2 (1 - \Omega_{*}) \sinh^2{\biggl[\sqrt{\frac{\beta}{2}} \frac{(\varphi - \varphi_{0})}{(1 - \Omega_{*}) \overline{M}_{\mathrm{P}}}\biggr]}.
\label{SOL4}
\end{equation}
Equations (\ref{SOL1}) and (\ref{SOL2}) satisfy the solution of Eq. (\ref{I1}). For $\beta H_{*} t < 1$ the solution is decelerated and from Eq. (\ref{SOL1}) we have $a(t) \simeq a_{*} ( \beta H_{*} t)^{1/\beta}$ where $H_{i} = 2/[3 ( w+ 1) t_{i}]$.
In the opposite limit (i.e. $\beta H_{*} t \gg 1$) the solution is accelerated with $H(t) \simeq H_{*}$ since 
$H(t) = H_{*}/[\tanh{(\beta H_{*} t)}]$  and $\dot{H} = - \beta H_{*}^2/[\sinh^2{(\beta H_{*} t)}]$.
\end{appendix}


\newpage
\begin{thebibliography}{99}

\bibitem{lich} A. Lichnerowicz, {\it Magnetohydrodynamics: waves and shock waves in curved space-time}, 
(Kluwer academic publisher, Dordrecht, 1994).

\bibitem{duality1}  S.~Deser and C.~Teitelboim,  Phys.\ Rev.\  D {\bf 13}, 1592 (1976).

\bibitem{duality2} S. Deser,  J. Phys. A {\bf 15}, 1053 (1982).

\bibitem{parker1} L.~Parker,  Phys.\ Rev.\ Lett.\  {\bf 21},  562 (1968).

\bibitem{birrell} N. D. Birrell and P. C. W. Davies, {\it Quantum fields in curved space}, 
(Cambridge University Press, Cambridge, UK, 1982).

\bibitem{parker2} L. Parker and D. Toms, {\it Quantum Field Theory in Curved Spacetime: Quantized Fields and Gravity}, (Cambridge University Press, Cambridge, UK, 2009).

\bibitem{rev1}  K. Enqvist, Int.\ J.\ Mod.\ Phys.\  D  {\bf 7}, 331 (1998).

\bibitem{rev2}  M.~Giovannini,  Int.\ J.\ Mod.\ Phys.\  D {\bf 13}, 391 (2004).

\bibitem{rev3} J.~D.~Barrow, R.~Maartens and C.~G.~Tsagas,  Phys.\ Rept.\  {\bf 449}, 131 (2007).

\bibitem{mgenesis} M. Giovannini,  Phys. Rev. D {\bf 62}, 123505 (2000).

\bibitem{b1}  H. Alfv\'en and C.-G. F\"althammer, {\it Cosmical Electrodynamics}, 2nd edn., (Clarendon press, Oxford, 1963).

\bibitem{b2} E. N. Parker, {\it Cosmical Magnetic Fields} (Clarendon Press, Oxford, 1979).

\bibitem{b3} Ya. B. Zeldovich, A. A. Ruzmaikin, D.D. Sokoloff  {\it Magnetic Fields in Astrophysics} (Gordon  Breach Science, New York, 
1983).

\bibitem{weyl} M.~Giovannini, Phys. Rev. D {\bf 85}, 101301(R) (2012).

\bibitem{m2} M.~Giovannini,  Phys.\ Rev.\ D {\bf 83}, 023515 (2011).

\bibitem{LL} A.~R. Liddle and S.~M. Leach,   Phys.\ Rev.\  D {\bf 68}, 103503 (2003).

\bibitem{nonvac1} P.~D.~B.~Collins and R.~F.~Langbein,ÊÊPhys.\ Rev.\ D {\bf 45}, 3429 (1992).

\bibitem{nonvac1a} I.~Sokolov,ÊClass.\ Quant.\ Grav.\  {\bf 9}, L61 (1992).

\bibitem{nonvac1b} M.~Gasperini, M.~Giovannini, G.~Veneziano,  Phys.\ Rev.\  D {\bf 48}, 439 (1993).
 
\bibitem{nonvac2} K.~Bhattacharya, S.~Mohanty and R.~Rangarajan, Phys.\ Rev.\ Lett.\  {\bf 96}, 121302 (2006).

\bibitem{nonvac3} W.~Zhao, D.~Baskaran and P.~Coles, Phys.\ Lett.\ B {\bf 680}, 411 (2009).
 
\bibitem{nonvac3a} I.~Agullo and L.~Parker,ÊPhys.\ Rev.\ D {\bf 83}, 063526 (2011).

\bibitem{nonvac3b} S.~Kundu, JCAP {\bf 1202}, 005 (2012).

\bibitem{LB}  R. A. Lyttleton and H. Bondi, Proc. Roy. Soc. A {\bf 252}, 313 (1959).

\bibitem{w1} P. Olesen, Phys. Lett. B {\bf 398}, 321 (1997).

\bibitem{w2}  A.~Brandenburg, K. Enqvist and P. Olesen Phys. \ Rev.\ D {\bf 54}, 1291 (1996).

\bibitem{w3}  K.~Subramanian and J.~D.~Barrow, Phys.\ Rev.\ D {\bf 58}, 083502 (1998).

\bibitem{w4}  M.~Christensson, M.~Hindmarsh,  Phys.\ Rev.\  D {\bf 60}, 063001 (1999).

\bibitem{dirac}  P. A. M. Dirac, Ann. of Math. {\bf 37}, 657 (1936).

\bibitem{barut}  A. O. Barut and R. B. Haugen, Ann. Phys. {\bf 71}, 519 (1972).

\bibitem{rorlich}  T. Fulton, F. Rohrlich, and L. Witten, Rev. Mod. Phys. {\bf 34}, 412  (1962).

\bibitem{DT1} B. Ratra,  Astrophys.\, J.\, Lett.  {\bf 391}, L1 (1992).

\bibitem{DT2} M.~Gasperini, M.~Giovannini, and G.~Veneziano, Phys. Rev. Lett. {\bf 75}, 3796 (1995).
 
\bibitem{DT3} M.~Giovannini,  Phys.\ Rev.\  D {\bf 64}, 061301 (2001).
 
\bibitem{DT5} K.~Bamba and M.~Sasaki,  JCAP {\bf 02}, 030 (2007); K. Bamba JCAP {\bf 10}, 015 (2007).

\bibitem{DT5a} P.A.M. Dirac, Nature {\bf 139}, 323 (1937); Proc. R. Soc. London A {\bf 165}, 199  (1938).

\bibitem{DT5b} P. Jordan, Z. Phys. {\bf 157}, 112 (1959); E. Teller, Phys. Rev. {\bf 73}, 801 (1948).

\bibitem{DT6} M. Giovannini, Phys.\ Lett.\  B {\bf 659}, 661 (2008);  JCAP {\bf 1004}, 003 (2010).

\bibitem{DT6a} K. Bamba, Phys. Rev. D {\bf 75} 083516 (2007).

\bibitem{DT7a}  K.~Enqvist, R.~N.~Lerner, O.~Taanila and A.~Tranberg,  arXiv:1205.5446 [astro-ph.CO].

\bibitem{DT7b} K.~Enqvist, Prog.\ Theor.\ Phys.\ Suppl.\  {\bf 190}, 62 (2011); K.~Enqvist, S.~Nurmi, O.~Taanila and T.~Takahashi,  JCAP {\bf 1004}, 009 (2010).

\bibitem{bounce1} M. Giovannini,  Class.\ Quant.\ Grav.\  {\bf 21}, 4209 (2004);  Phys.\ Rev.\ D {\bf 70}, 103509 (2004);
M.~Gasperini, M.~Giovannini and G.~Veneziano,  Phys.\ Lett.\ B {\bf 569}, 113 (2003); Nucl.\ Phys.\ B {\bf 694}, 206 (2004).

\bibitem{bounce2} M. Novello and S.E. Perez Bergliaffa, Phys. Rep. {\bf 463}, 127 (2008).

\bibitem{wmap}  D.~N.~Spergel {\it et al.},  Astrophys.\ J.\ Suppl.\  {\bf 148}, 175 (2003); D.~N.~Spergel {\it et al.},
{\em ibid.} \ {\bf 170}, 377 (2007).

\bibitem{wmap7} C.~L.~Bennett {\it et al.},  Astrophys.\ J.\ Suppl.\  {\bf 192}, 17 (2011);   N.~Jarosik {\it et al.},  
Astrophys.\ J.\ Suppl.\  {\bf 192}, 14 (2011); J.~L.~Weiland {\it et al.},  Astrophys.\ J.\ Suppl.\  {\bf 192}, 19 (2011).

\bibitem{wmap7a} D.~Larson {\it et al.}, Astrophys.\ J.\ Suppl.\  {\bf 192}, 16 (2011); B.~Gold {\it et al.},  Astrophys.\ J.\ Suppl.\  {\bf 192}, 15 (2011);  
E.~Komatsu {\it et al.},   Astrophys.\ J.\ Suppl.\  {\bf 192}, 18 (2011).

\bibitem{yaglom} A. S. Monin and A. M. Yaglom, {\it Statistical Fluid Mechanics}, (Dover Publications, Mineola, New York, 1979).

\bibitem{landau2} L. D. Landau and E. M. Lifshitz, {\it Fluid Mechanics}, (Pergamon Press, Oxford, 1987).

\bibitem{compress} M.~Giovannini, Phys.\ Rev.\ D {\bf 85}, 043006 (2012).

\bibitem{abr} M. Abramowitz and I. A. Stegun, {\it Handbook of Mathematical Functions} (Dover, New York, 1972).

\bibitem{tric} A. Erdelyi, W. Magnus, F. Obehettinger, and F. Tricomi, {\it Higher Trascendental Functions} (Mc Graw-Hill, New York, 1953).  

\bibitem{loudon2}  R. Loudon, {\it The quantum theory of light} (Clarendon Press, Oxford, 1983).

\bibitem{mandel}  L. Mandel and E. Wolf, {\it Optical coherence and quantum optics}, (Cambridge University Press, Cambridge, 1995).

\bibitem{CC1} J.~Ahonen, Phys.\ Rev.\ D {\bf 59}, 023004 (1999).

\bibitem{CC2} J.~Ahonen and K.~Enqvist,  Phys.\ Lett.\ B {\bf 382}, 40 (1996).

\bibitem{CC3} H. Heiselberg, Phys. Rev. D {\bf 49}, 4739 (1994).

\bibitem{lid} M.~Giovannini, Phys.\ Rev.\  D {\bf 60}, 123511 (1999); Phys.\ Lett.\  B {\bf668}, 44 (2008).

\bibitem{KT} E. M. Lifshitz and L. P. Pitaevskii, {\it Physical Kinetics} (Pergamon, Oxford, England, 1980)

\bibitem{KTA}  J. Bernstein, {\it Kinetic theory in the expanding universe} (Cambridge Univ. Press, Cambridge, England, 1988).

\bibitem{olesmagn1}  P.~Olesen, Phys.\ Lett.\ B {\bf 366}, 117 (1996).
  
\bibitem{olesmagn2}   O.~Coceal, W.~A.~Sabra and S.~Thomas, Phys.\ Lett.\ B {\bf 389}, 655 (1996).

\bibitem{mxpr} M. Giovannini, Class. Quantum Grav. {\bf 29}, 155003 (2012).

\end{thebibliography}
\end{document}